\documentclass[twocolumn,floatfix]{aastex631}

\usepackage{bm}
\usepackage{upgreek}
\usepackage{float}
\graphicspath{ {./Images/} }

\newcommand{\msun}{\,\textrm{M}_\odot}
\newcommand{\vpol}{v_{\rm pol}}
\newcommand{\tps}{t_{\rm ps}}
\newcommand{\Peff}{\Phi_{\rm eff}}
\newcommand{\Omzsq}{\Omega_0^2}
\newcommand{\Omksq}{\Omega_{\rm K0}^2}
\newcommand{\rcent}{r_{\rm c}}
\def\alfven{Alfv\'en }
\def\alfvenic{Alfv\'enic }

\begin{document}

\title{Outflows Driven from a Magnetic Pseudodisk}
\journalinfo{The Astrophysical Journal 2024}

\author[0000-0003-0855-350X]{Shantanu Basu}
\affiliation{Department of Physics and Astronomy,
University of Western Ontario,
London, Ontario N6A 3K7, Canada}
\affiliation{Institute for Earth and Space Exploration, University of Western Ontario, London, Ontario N6A 5B7, Canada}
\affiliation{Department of Earth and Planetary Sciences, Faculty of Sciences, Kyushu University, Fukuoka  819-0395, Japan}
\author[0000-0002-2012-2609]{Mahmoud Sharkawi}
\affiliation{College of Engineering, Al Ain University, Abu Dhabi, United Arab Emirates}
\affiliation{Department of Mathematics,
University of Western Ontario,
London, Ontario N6A 5B7, Canada}
\author[0000-0002-0963-0872]{Masahiro N. Machida}
\affiliation{Department of Earth and Planetary Sciences, Faculty of Sciences, Kyushu University, Fukuoka  819-0395, Japan}
\affiliation{Department of Physics and Astronomy,
University of Western Ontario,
London, Ontario N6A 3K7, Canada}

\begin{abstract}
Outflows play a pivotal role in star formation as one of its most visible markers and a means of transporting mass, momentum, and angular momentum from the infalling gas into the surrounding molecular cloud. Their wide reach (at least  thousands of au) is a contrast to typical disk sizes ($\sim 10-100$ au). We employ high-resolution three-dimensional nested-grid nonideal magnetohydrodynamic (MHD) simulations to study outflow properties in the Class 0 phase. We find that no disk wind is driven from the extended centrifugal disk that has weak magnetic coupling. The low-velocity winds emerge instead from the infalling magnetic pseudodisk. Much of the disk actually experiences an infall of matter rather than outflowing gas. Some of the pseudodisk wind (PD-wind) moves inward to regions above the disk and either falls onto the disk or proceeds upward. The upward flow gives the impression of a disk wind above a certain height even if the gas is originally emerging from the pseudodisk. The PD-wind has the strongest flow coming from a disk interaction zone that lies just outside the disk and is an interface between the inwardly advected magnetic field of the pseudodisk and the outwardly diffusing magnetic field of the disk. The low-velocity wind exhibits the features of a flow driven by the magnetic pressure gradient force in some regions and those of a magnetocentrifugal wind in other regions. We interpret the structure and dynamics of the outflow zone in terms of the basic physics of gravity, angular momentum, magnetic fields, and nonideal MHD. 

\end{abstract}


\section{Introduction} \label{sec:intro}

The magnetic field plays a vital role in gravitational collapse that leads to star formation.
Infalling matter is channeled by the magnetic field into a flattened pseudodisk, and the magnetic field can extract angular momentum from the rotating gas and launch outflows. These influences of the magnetic field act to regulate the formation of a centrifugally-supported disk. Building a self-consistent picture of the magnetized collapse of a star-forming core through to the formation of a star-disk-outflow system is a key goal in modern astrophysics.


Outgoing flows of material from protostar-disk systems have typically been categorized into two types: relatively low velocity ($ \lesssim 10-50$ km s$^{-1}$) outflows that are characterized by molecular (typically CO) line emission and higher velocity ($ \sim 100-500$ km s$^{-1}$) flows (or jets) typically seen in optical or near-infrared emission but more recently in molecular lines as well \citep[e.g., see the review papers by][]{bal16,lee20,tsu23,pas23}. 
In spite of their characteristically high speeds, the jets do not meaningfully contribute to the dynamical evolution of the protostellar core, whereas low-velocity outflows expel a significant portion of the infalling matter and may determine the final stellar mass \citep[see, e.g.,][]{pas23}.
While the effect of the outflows can be seen at vertical distances $\leq 1000$ au, it remains difficult to determine their actual driving points. 

The magnetic field seems to be key to understanding the outflow phenomenon, and simulations have to resolve the motions parallel to the magnetic axis in order to capture the effect. \cite{tom98,tom02} used two-dimensional ideal magnetohydrodynamic (MHD) simulations in cylindrical coordinates to show outflow launching as a self-consistent outcome of protostellar collapse. The nonideal MHD effects are also crucial to model the full time evolution and identify the launch regions of the outflows. Using three-dimensional resistive MHD numerical simulations, \citet{mac06,mac08} found that low-velocity flows are driven from near the first adiabatic core surface (at radius $\sim 10$ au), while an inner high-velocity jet is driven from near the inner protostar (second core).
Subsequent three-dimensional nonideal MHD simulations that resolve the second core have revealed a qualitatively similar picture \citep[e.g.,][]{tom15,tsu15,wur18}, although there are quantitative differences that may be attributed to differences in the initial conditions, numerical techniques, and resolution. These simulations that resolve the stellar core generally terminate within a few hundred years after the second core formation, so they cannot model the subsequent phases that are actually observed. The model of this type with the longest protostellar duration \citep[2000 yr;][]{mac19}, shows the existence of up to three distinguishable velocity components at this early time, including a low-velocity outflow at the disk edge and a high-velocity jet from near the protostar. The magnetocentrifugal wind (MCW) mechanism, which operates in magnetically-dominated regions \citep{bla82}, likely plays an important role in the launch of the low-velocity winds. The inner jet is driven by the magnetic pressure gradient force (MPGF) perpendicular to the disk \citep{uch85} in a system where the rotation and magnetic axes are aligned, and even when these axes are somewhat misaligned \citep{machida2020,hir20}. An alternative mechanism for the inner jet is the interaction of a stellar magnetic field with the inwardly advected disk magnetic field, resulting in an MCW along open field lines of the disk that are very close to the protostar \citep{shu94}.

To properly compare simulations with observations, it is necessary for the simulations to reach the observable Class 0/I phases of protostellar age $\sim 10^4-5\times 10^5$ yr. For this purpose, a central sink cell is typically employed in order to keep the time steps from becoming too small to allow long-term integration. 
Three-dimensional nonideal MHD simulations that employ a central sink cell \citep[e.g.,][]{mac13} have shown that the outflow driving continues to occur from the disk-like region surrounding the protostar as it continues to grow in size over time. 
In this paper we employ a three-dimensional nonideal nested-grid MHD simulation with a central sink cell and follow the evolution into the Class 0 protostellar phase. We explore the physics of the launch region of the low-velocity wind, including the disk, the pseudodisk, and the interaction zone between them.

For our purposes we describe the low-velocity wind (or outflowing gas) of maximum speed several km\,s$^{-1}$ as an ``outflow.''
Thus, such flows may not correspond to a powerful outflow such as that seen in L1551, for which \citet{snell80} first observed the (protostellar) outflow in CO line emission.
However, recent ALMA observations have confirmed that not all young stellar (or Class 0/I) objects exhibit such a powerful outflow.
The weak or low-velocity outgoing flow with maximum velocity $<10$\,km\,s$^{-1}$ is also called an outflow in observations \citep[e.g.,][]{aso19,sato23}. 
In the following, for convenience, we refer to an outgoing flow that appears in our simulations as an outflow. 



\section{Methods}
We employ the three-dimensional resistive MHD nested-grid code used by \cite{mac08} and \cite{mac13} with many of the same initial settings. 
The following resistive MHD equations are solved, which include self-gravity:
\begin{eqnarray}
\frac{\partial \rho}{\partial t} + \nabla \cdot (\rho\, \bm{v}) & = & 0 \, , \\
\rho\, \frac{\partial \bm{v}}{\partial t} + \rho\, (\bm{v} \cdot \nabla)\,\bm{v} & = & -\nabla P + \frac{\bm{j}}{c} \times \bm{B} - \rho\,  \nabla \Phi \, , \\
\frac{\partial \bm{B}}{\partial t} - \nabla \times (\bm{v} \times \bm{B}) & = &  \eta\, \nabla^2 \bm{B} \, , \label{eq:induction}\\
\nabla^2 \Phi & = & 4\pi G \rho \, . 
\end{eqnarray}
Here, $\rho, \bm{v}, P, \bm{B}, \eta$, and $\Phi$ are the density, velocity, pressure, magnetic field, resistivity, and gravitational potential, respectively. The electric current density is $\bm{j}=c\,(\nabla \times \bm{B})/4\pi$. 
Equation (\ref{eq:induction}) is an approximation used in this work. The more correct form of the induction equation has the ohmic resistivity term as $\nabla \times (-\eta\, \nabla \times \bm{B})$ \citep[see][]{mac18,machidabasu2020}. See Appendix~\ref{ap:res} for an example case of one model that uses the correct induction equation and reaches similar results.

A Bonnor-Ebert density profile is used to construct the initial state with central number density and isothermal temperature given by $n_c = 6 \times 10^5 \, \textrm{cm}^{-3}$ and $T = 10$~K, respectively. The initial core mass and radius are $M_{\rm cl} = 2\msun$ and $R_{\rm cl} = 1.2 \times 10^4$ au, respectively. The initial cloud density is amplified by a factor $f = 1.68$ to promote contraction \citep[see][]{mac20}. The initial ratio of the thermal energy to the absolute value of gravitational energy is $\alpha_0 = 0.42$. We study a set of four models of different initial uniform magnetic field strength $B_0$ and rigid rotation rate $\Omega_0$. These, in turn, imply different values of the mass-to-flux ratio $\mu_0$ (normalized by the critical value $1/(2\pi \sqrt{G})$) and ratio of rotational to gravitational energy $\beta_0$, respectively. Parameters for the different models are shown in Table~\ref{table:models}.
\begin{deluxetable}{cccccc}[t]
\tablecaption{Parameters of Initial Clouds}
\tablecolumns{6}
\tablenum{1}
\tablewidth{0pt}
\label{table:models}
\tablehead{
  \colhead{Model} &
  \colhead{$B_0$ ($\upmu$G)} &
  \colhead{$\Omega_0$ (s$^{-1}$)} &
  \colhead{$\mu_0$} &
  \colhead{$\beta_0$} &
  \colhead{$\alpha_0$} 
}
\startdata
B01 & $32$  & $1.2\mathrm{e}{-13}$ & 2   & 0.0224 & 0.42 \\
B02 & $16$  & $1.2\mathrm{e}{-13}$  & 4  & 0.0224 & 0.42 \\
R01 & $32$  & $8.3\mathrm{e}{-14}$ & 2   & 0.01 & 0.42 \\
R02 & $32$  & $1.4\mathrm{e}{-13}$ & 2   & 0.03 & 0.42 \\
\enddata
\tablecomments{
Column (1): model name.
Column (2): initial magnetic field strength. 
Column (3): initial rotation rate. 
Column (4): initial mass-to-flux ratio. 
Column (5): initial ratio of rotational to gravitational energy. 
Column (6): initial ratio of thermal to gravitational energy. 
}
\end{deluxetable}

The ohmic dissipation is modeled by the effective resistivity $\eta$ formulated in \cite{mac07,mac08} and taken from \cite{nak02}. It is estimated to be a function of number density and temperature \citep[see also][]{OlivaKuiper2023a}. In the collapsing core, it is assumed that protostar formation occurs when the number density $n > n_{\rm cr} = 10^{12} \, \textrm{cm}^{-3}$ at the cloud center. A sink cell at the center of the computational domain is adopted, with radius $r_{\rm sink} = 2$ au. In the region $r < r_{\rm sink}$, gas having a number density of $n > n_{\rm cr}$ is removed from the computational domain and its mass is added to the protostar at each time step. The sink technique allows us to explore the long-term evolution of star-forming clouds and determine the final stellar mass as well as the morphological structure of the protostellar disk \citep[e.g.,][]{tom17}. Nonetheless, the structure inside the sink cell itself is not resolved in such simulations. Additionally, the magnetic flux is removed via ohmic dissipation inside the sink. For $r > r_{\rm sink}$, 
we close the system of equations with a barotropic relation
\begin{equation}
 P = c_{\rm s}^2 \rho \left[ 1 + \left( \frac{\rho}{\rho_{\rm cr}}\right)^{\gamma-1} \right]   \, ,
\end{equation}
where $c_{\rm s}$ is the isothermal sound speed at the initial cloud temperature, $\gamma = 7/5$ is the adiabatic index, and $\rho_{\rm cr} = 1.92 \times 10^{-14}$ g~cm$^{-3}$ is the critical mass density.

To ensure coverage of a large dynamic range of scales, 
these equations are solved on a nested grid \citep{mat03,mac05a,mac05b}.
The nested grid contains 13 different sized grids indexed by $L=1-13$; each increasing index value corresponds to a finer grid with a halved grid span and cell width (for a schematic view of the grid see fig. 1 of \citealt{mat03}).
Each grid has different grid span and cell width, but the same number of cells $(i,j,k) = (64,64,64)$.

The grid span and cell width of the first level ($L=1$) are 
$3.7 \times 10^5$ and $5.9 \times 10^3$\,au, respectively, while those for the finest grid ($L=13$)  are 92 and 1.44\,au, respectively.
The initial molecular (or prestellar) cloud core is immersed in the $L=5$ grid, outside which a uniform lower density medium is imposed.
The large area outside the prestellar cloud core ($L=1-4$) is utilized in order to suppress artificial reflection of Alfv\'en waves \citep{mac13}. 
The calculation starts with five levels of the grid ($L=1-5$) and a finer grid is automatically generated so as to resolve the Jeans wavelength with at least 16 cells. 
We simulate the core evolution
with levels ranging from $L = 1$ to $L = 13$.




Many steady-state theories of MHD outflows and models of ideal and nonideal MHD collapse have been built in the two-dimensional axisymmetric approximation. Hence, we feel it is particularly important to study our three-dimensional and non-steady-state model in a way that can be linked with physical concepts derived separately in past works. Therefore, in this paper we present our simulation results primarily in the form of two-dimensional cuts in the $x-z$ plane or by azimuthally-averaged quantities.

\section{Results}
\label{sec:results}

\subsection{Overview}

\begin{figure*}[t]
\begin{center}
\includegraphics[width=1.0\linewidth]{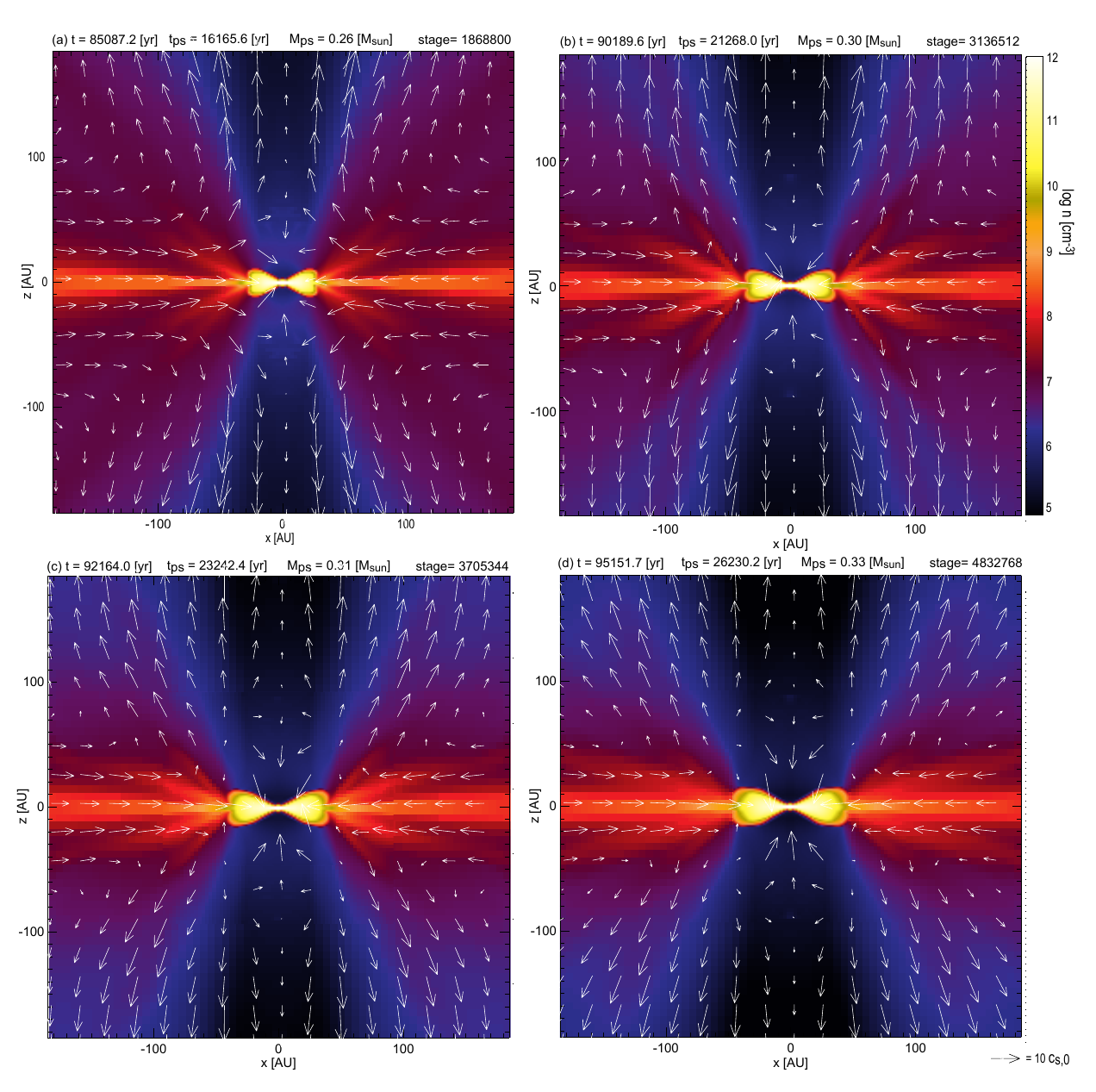}
\end{center}
\caption{Cuts along the $x-z$ plane of the number density with overlaid velocity vectors for model B01. Four different snapshots are shown during the protostellar phase. The sequence goes from top left to top right, then bottom left to bottom right. The times correspond to $t=[85.1,90.2,92.2,95.2]$ kyr, or $t_{\rm ps} = [16.2,21.3,23.2,26.2]$ kyr since protostar formation. The color table at top right shows the number density values and the vectors are normalized to $10\,c_{\rm s}$ as shown at the bottom right.}
\label{fig:timesequence}
\end{figure*}

We study a range of models with different physics in order to reach robust conclusions about disk and outflow formation in the Class 0 phase. After presenting general trends of the various models, we focus in detail on one fiducial model.

Figure~\ref{fig:timesequence} shows a time sequence of two-dimensional cuts in the $x-z$ plane of model B01 shown at level $L=11$. These evolutionary stages, all of which are in the protostellar Class 0 phase, show a well-developed disk and outflowing gas. The outflowing gas is steady during these final $\simeq 10$ kyr of the simulation. The disk radius is growing with time, and the disk density image has a flared structure since the vertical scale height is greater in regions of weaker gravity. The pattern shows a remarkable structure, with gas with a modest velocity ($v_z>0$) emerging from a region outside the disk and reaching large heights without strong interactions with the outer environment. The region immediately above the disk shows inflowing rather than outflowing gas. The inflow along the equatorial plane and the outflow motions are all supersonic at these distances ($\lesssim 200$ au from the protostar). We investigate the quantitative details of the final snapshot (bottom right panel of Figure~\ref{fig:timesequence}) in the following subsection.

\begin{figure*}[t]
\begin{center}
\includegraphics[width=1.0\linewidth]{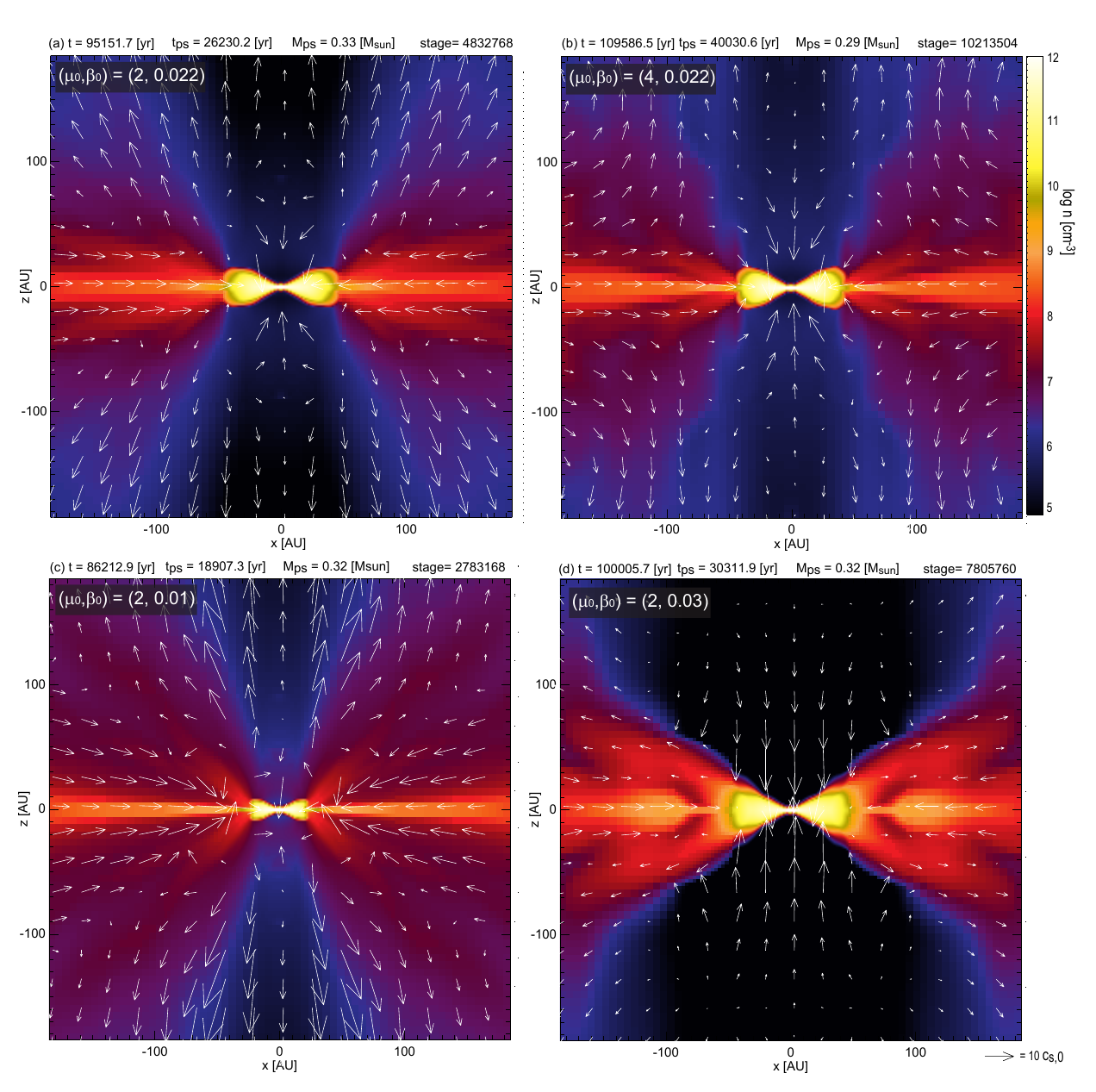}
\end{center}
\caption{Same as Figure~\ref{fig:timesequence} but for the four different models listed in Table~\ref{table:models}. The different models are shown at approximately the same stage of growth of the central mass $M_*$. Top left: model B01, $M_*=0.33\msun,\,t=95.2\;{\rm kyr},\,t_{\rm ps}=26.2\;{\rm kyr}$.
Top right: model B02, $M_*=0.29\msun,\,t=110 \;{\rm kyr},\,t_{\rm ps}=40.0\;{\rm kyr}$.
Bottom left: model R01, $M_*=0.32\msun,\,t=86.2\;{\rm kyr},\,t_{\rm ps}=18.9\;{\rm kyr}$.
Bottom right: model R02, $M_*=0.32\msun,\,t=100\;{\rm kyr},\,t_{\rm ps}=30.3\;{\rm kyr}$.
}
\label{fig:parameters}
\end{figure*}

Figure~\ref{fig:parameters} shows a snapshot at approximately the same evolutionary time (measured by the protostar mass $M_*$) of the four models listed in Table~\ref{table:models}. The fiducial model B01 is shown in the top left, and this panel is the same as the bottom right panel of Figure~\ref{fig:timesequence}. The top right panel shows model B02, which has a weaker initial magnetic field strength, hence $\mu_0=4$, but otherwise the same initial parameters as model B01. The disk size in model B02 is essentially the same size as in model B01. This means that the angular momentum budget of the infalling pseudodisk is similar in both models. The model B02 does show a larger region of infall above the disk than does model B01. This is due to the weaker magnetic field driving of the outflow. We return to these topics of magnetic braking and outflow driving in our later detailed examination of model B01.
The bottom left panel shows model R01, which differs from model B01 only in having a lower initial rotation rate, hence $\beta_0=0.01$. In this case, the reduced angular momentum budget leads to a disk about half as large as in model B01. This is consistent with the scaling $\beta \propto J^2/(M^3R)$ for a rigidly rotating sphere of angular momentum $J$, mass $M$, and radius $R$. In that case, if collapse occurs from $\beta=\beta_0$ until $\beta=1$ with mass and angular momentum conservation, then the ratio of final to initial radius $R/R_0 \propto \beta_0$. The region of infall above the disk has approximately the same vertical extent as in model B01, supporting the idea that this feature is dependent on the magnetic field strength and not the rotation rate. However, the smaller disk and consequent smaller launch radius of the outflow means that the flow launches at a greater escape speed. The bottom right panel shows model R02, which has $\beta_0=0.03$. The disk size is larger than in model B01 by about a factor 1.5, as expected based on the scaling discussed above. Compared to model B01, the larger (smaller) disk size in model R02 (R01) pushes the wind launch region to larger (smaller) radii and creates a higher (lower) region of infalling gas above the disk and lower (greater) outflow speeds.

\subsection{Fiducial Model}

Having identified some general trends in the models, which share similar qualitative features, we now focus in detail on the fiducial model B01. 
The model is run until the central mass has reached a value $M_*=0.33\,\msun$, corresponding to an evolutionary time $t=95.2$ kyr, which is a time $\tps = 26.2$ kyr since the formation of the protostar. Our goal is to understand properties of the disk, pseudodisk, and outflow zone during the accretion phase in what would be regarded observationally as a Class 0 protostar.

\begin{figure}
\plotone{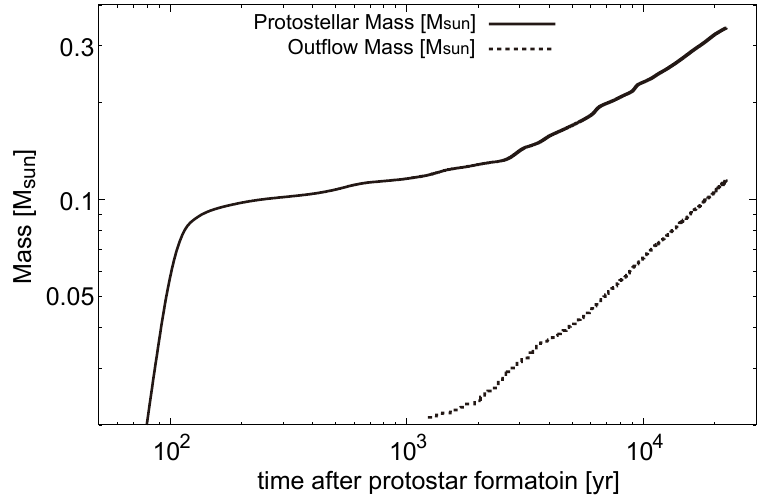}
\caption{Time evolution of the mass in the central protostar and in the outflow for model B01.}
\label{fig:masses}
\end{figure}

Figure~\ref{fig:masses} shows the time evolution of the central protostar mass and the total mass of the outflowing gas. Times are measured since the formation of the central protostar. After an initial adjustment period, there is an approximately linear growth of the masses after a time $\tps \simeq 3$ kyr. At the end of the simulation at $\tps = 26.2$ kyr, the protostar and outflow mass are $0.33\,\msun$ and $0.11\,\msun$, respectively.

We analyze the spatial structure of various physical quantities in subsequent plots, mostly at the final snapshot. All subsequent analysis is done after transforming the Cartesian nested grid values into cylindrical coordinates ($r,\phi,z$) and taking an azimuthal average for the ($r,z$) spatial plots.

\begin{figure}
\plotone{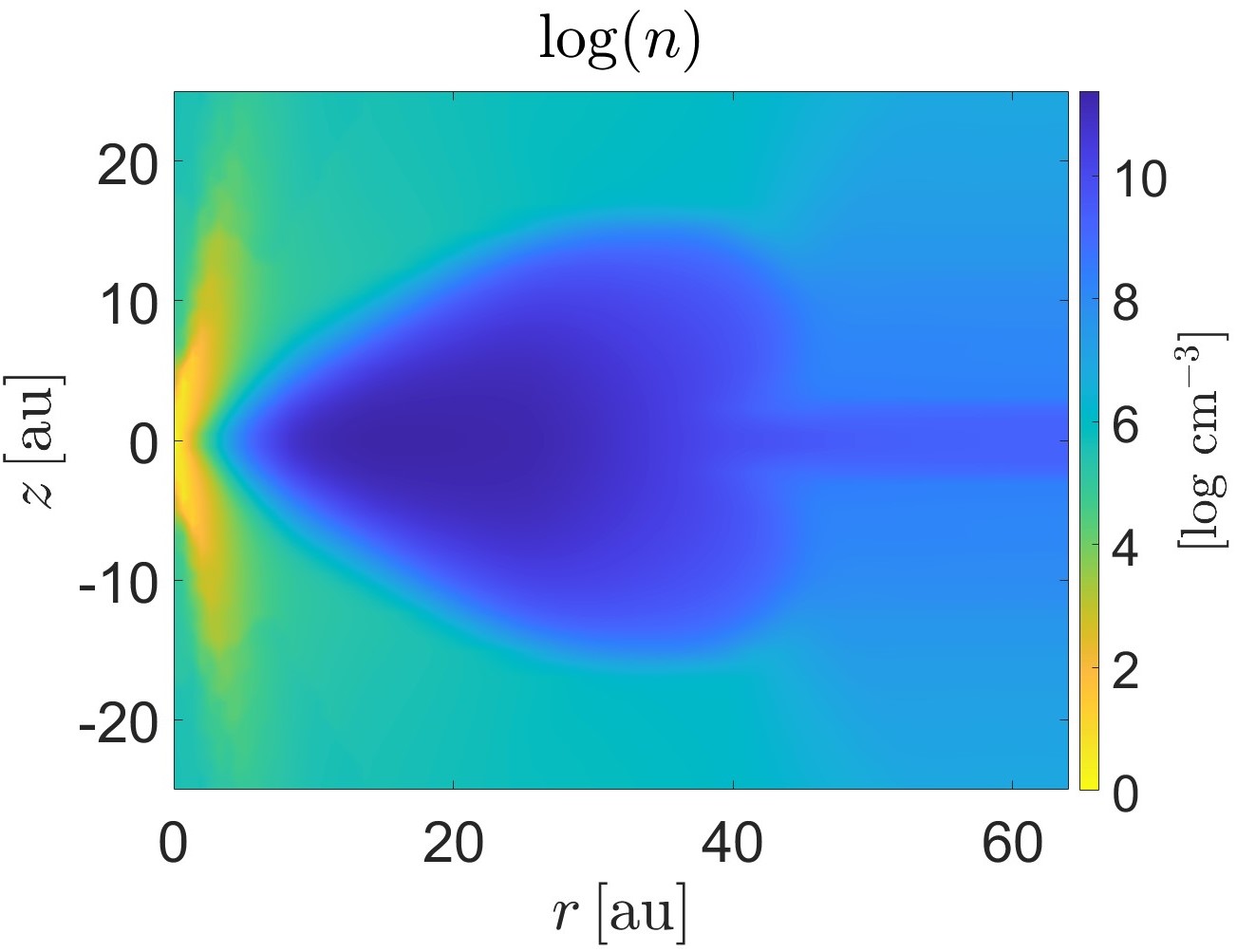}
\plotone{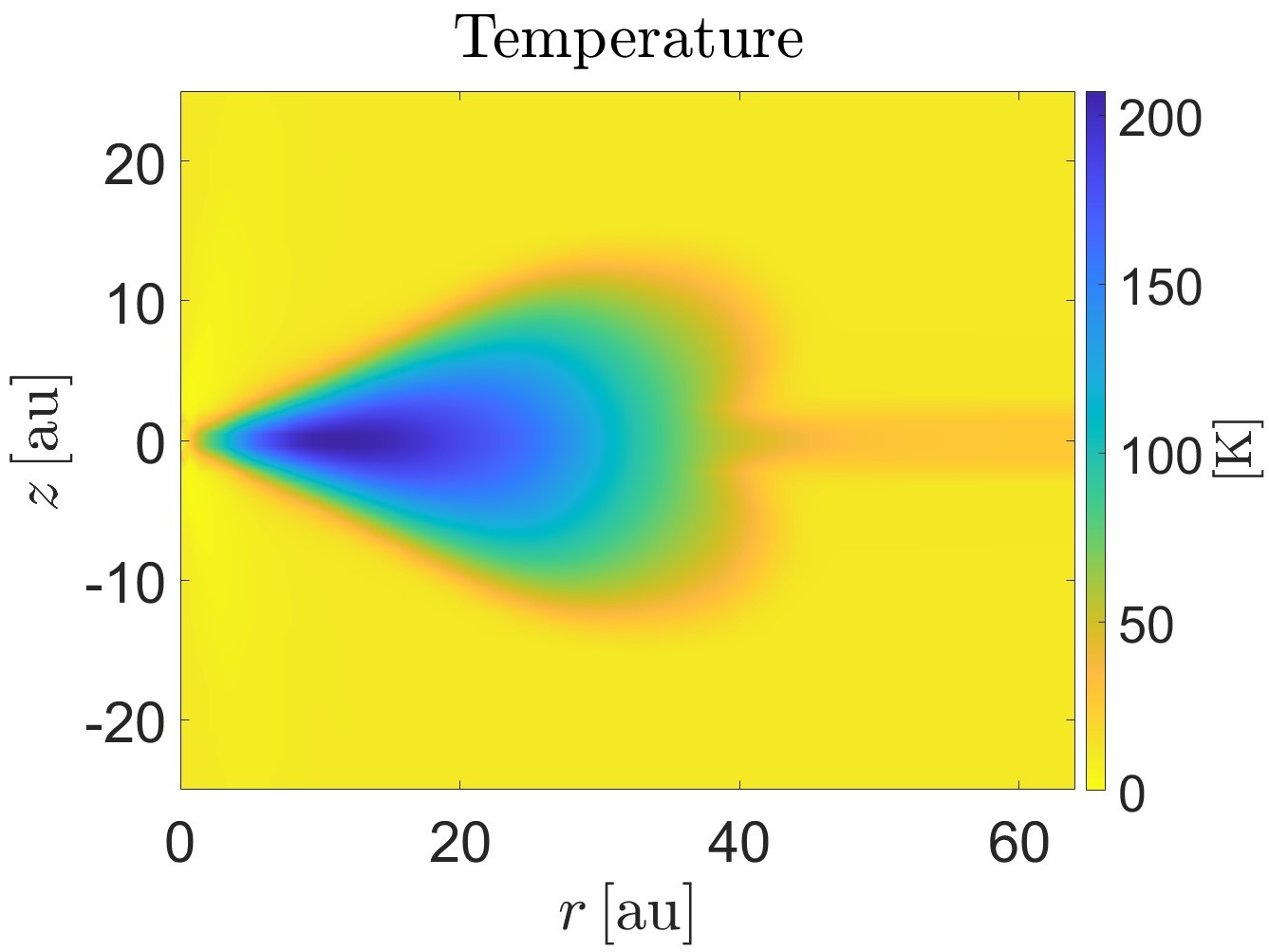}
\caption{Top: Number density in the innermost grid level $L=13$ with the color table in logarithmic values. Bottom: Temperature field for the same grid level and time as the top panel. The color table is in Kelvin (K).}
\label{fig:densityimage}
\end{figure}

Figure~\ref{fig:densityimage} shows an image of the number density at the highest refinement level $L=13$. A centrifugally-supported disk of size $\simeq 40$ au is clearly visible and is flared due to the decreasing vertical gravity component $g_z$ as radius increases. The gravity of the protostar dominates the self-gravity of the core at these radii. Outside the disk, a thin high-density distribution is seen, corresponding to the inner region of the pseudodisk that is formed by flattening along the direction of the background magnetic field. The pseudodisk is dynamically infalling but builds up a rapid rotation in its inner regions, which we discuss later in the paper. The bottom panel of Figure~\ref{fig:densityimage} shows the temperature field in the same region shown in the top panel. The disk has significantly greater temperature than the pseudodisk, which partially explains its greater scale height (see Section~\ref{sec:discussion} for further discussion on this). The gravity in the disk region is dominated by that of the central protostar of mass $M_*$, and a decrease with radius of the vertical gravitational field $g_z \simeq GM_*z/r^3$ for $z \ll r$ leads to an increasing disk scale height versus radius.

\begin{figure}
\epsscale{1.1}
\plotone{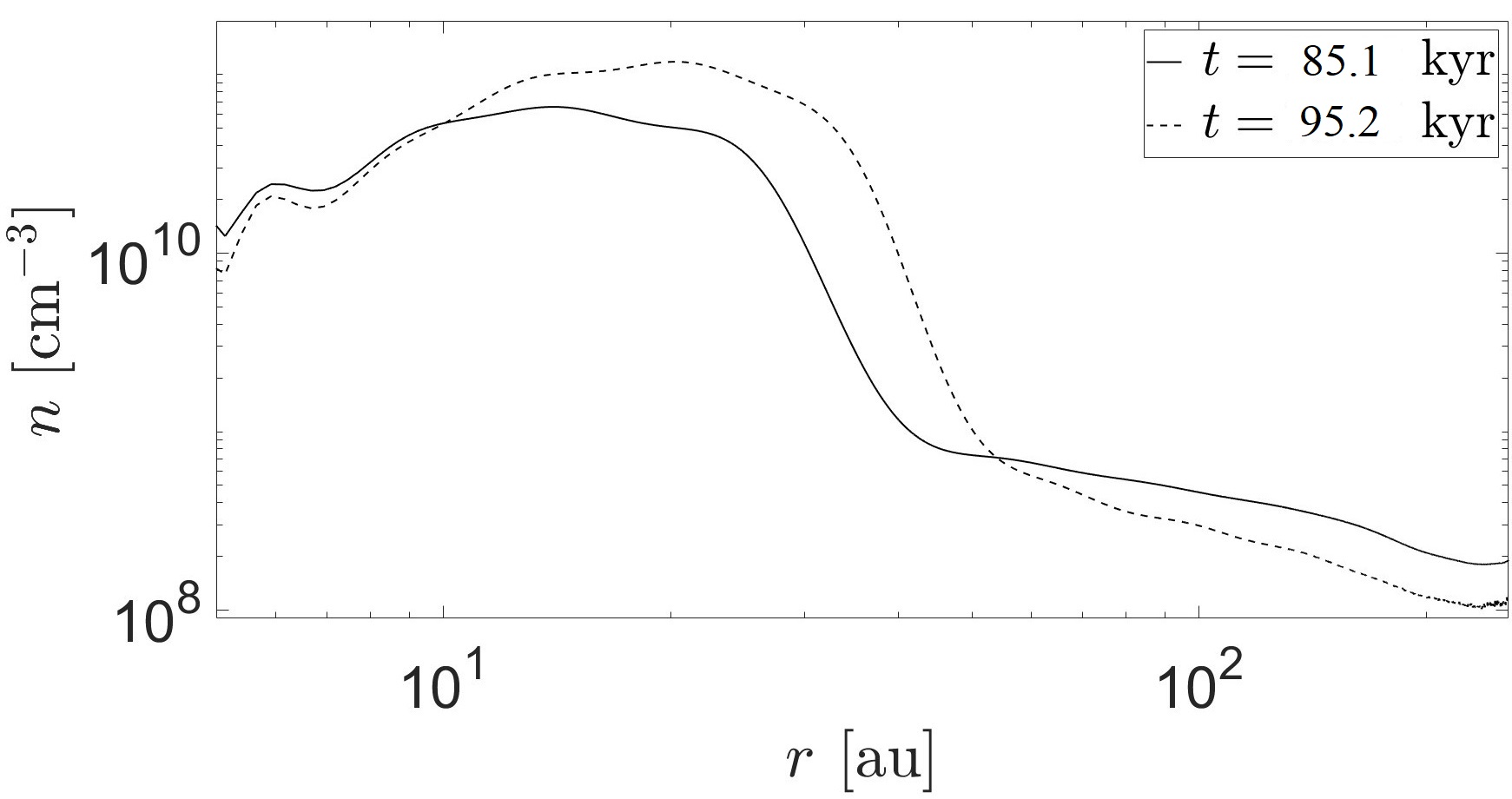}
\caption{Number density along the equatorial plane $z=0$ at level $L=11$. Two evolutionary times are shown.}
\label{fig:densitylinegraphs}
\end{figure}

Figure~\ref{fig:densitylinegraphs} shows the density profile along the equatorial plane $z=0$ at an inner scale ($L=11$) of the simulation and at two evolutionary times. The disk has a relatively sharp edge in density, and its location moves outward in the $\simeq 10$ kyr time interval between the snapshots. The rapid spatial increase of density inside the disk leads to a high resistivity, hence the nonideal MHD effects dominate inside the disk, as will be seen later. Previous work \citep{mac11} has established that the first adiabatic core of size $\sim 1-10$ au is the place where the ohmic dissipation first dominates, and that the first core forms the centrifugal disk as it collapses. Subsequent disk growth occurs as infalling mass shells encounter the disk, creating increasingly larger regions of high density and centrifugal support. The magnetic braking catastrophe \citep{all03,gal06} is averted by the nonideal MHD processes that initially start on the $\sim 1-10$ au scale \citep{dap12,tom17} and move outward as the disk is formed and grows in size \citep[see discussion in][]{tsu23}.


\begin{figure}
\epsscale{1.1}
\plotone{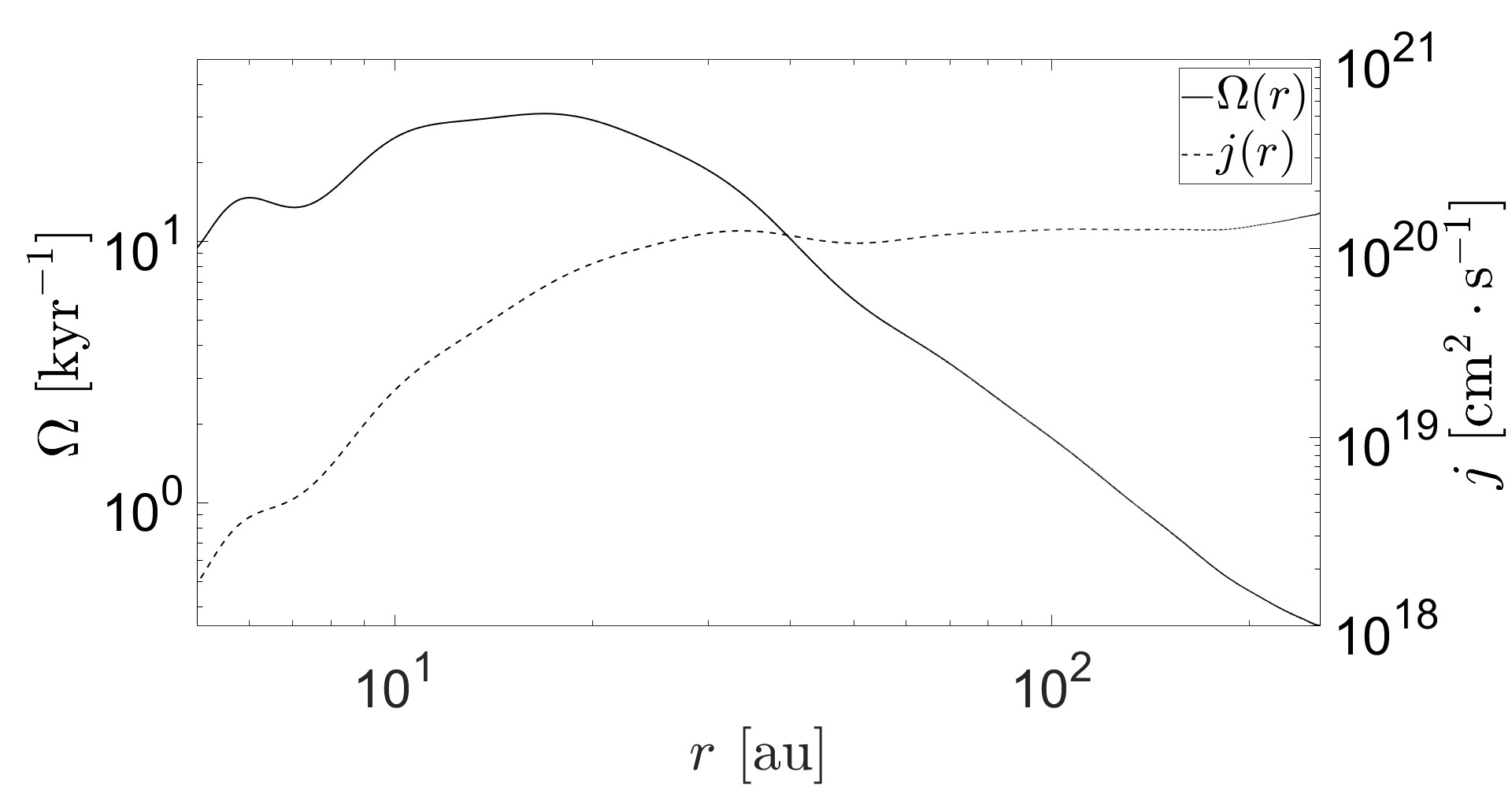}
\caption{Angular velocity $\Omega$ and specific angular momentum $j = \Omega\, r^2$ along the equatorial plane $z=0$ at level $L=11$ at a time $t=95.2$ kyr, or $\tps = 26.2$ kyr.}
\label{fig:rotation}
\end{figure}

\begin{figure*}[ht!]
\plotone{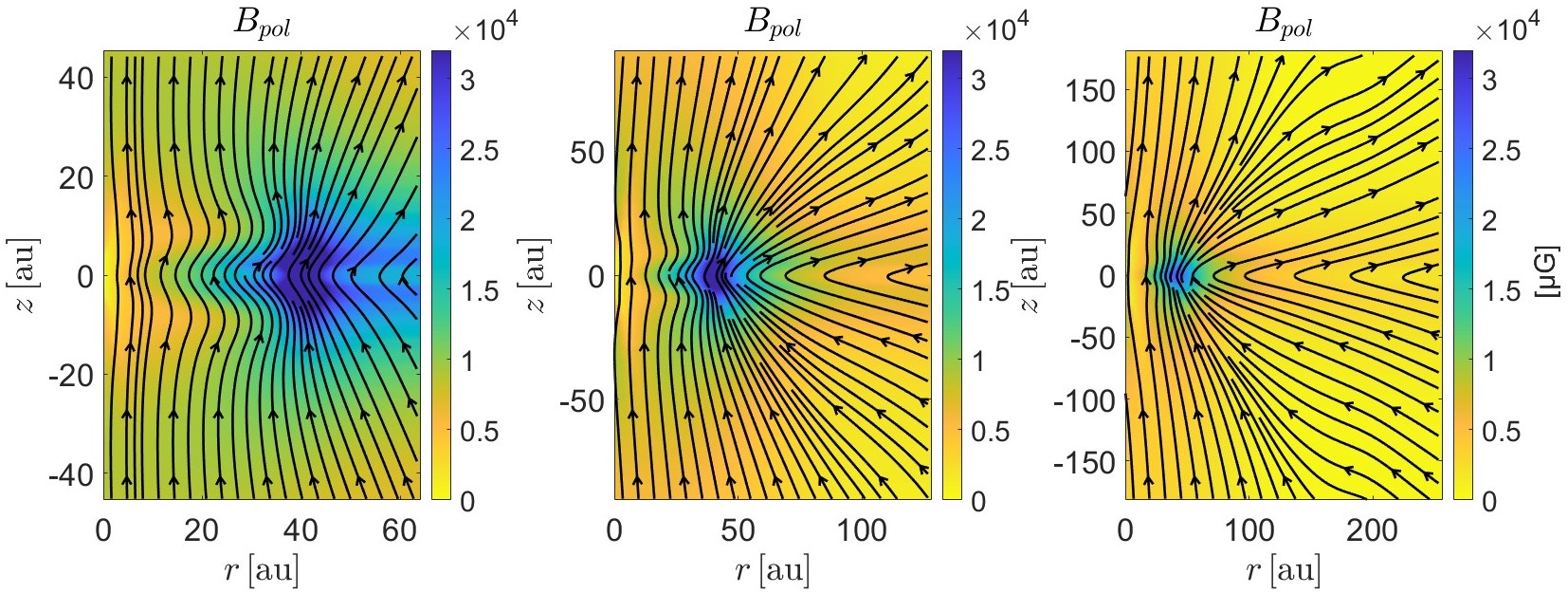}
\caption{Lines of poloidal magnetic field at levels $L=13,12,11$ with the color table representing the magnetic field strength in units of $\upmu \textrm{G}$.}
\label{fig:bzplots}
\end{figure*}

Figure~\ref{fig:rotation} shows the angular velocity $\Omega$ and specific angular momentum $j$ profiles along $z=0$ at level $L=11$. The innermost 10 au is affected by the presence of the central sink. The most important features of this plot are the breaks in the profiles at the edge of the disk at $\simeq 40$ au, and the profiles outside the disk in the infalling envelope (pseudodisk). The profiles are approximately consistent with a spatially constant $j$ and therefore $\Omega \propto r^{-2}$. This profile in the infalling envelope has been found in self-similar profiles of rotating hydrodynamic collapse \citep{ter84,sai98} as well as semi-analytic models of collapse with angular momentum conservation \citep{tak16,das22}. It has
also been found in thin-disk rotating magnetized collapse with magnetic braking and nonideal MHD effects \citep{dap10,dap12} since the braking time is longer than the collapse time in the pseudodisk region \citep[see fig. 13 of][]{dap12}. Our three-dimensional simulation that follows magnetic braking and outflows in the vertical direction gets this same overall result. The approximately constant value of $j \simeq 1.5 \times 10^{20}$ cm$^2$ s$^{-1}$ in the pseudodisk yields a centrifugal barrier $r_{\rm c} = j^2/GM_* \simeq 35$ au, using the current protostar mass $M_*=0.33\,\msun$, showing good consistency between our simulation result and an analytic estimate.

Figure~\ref{fig:bzplots} shows elements of the magnetic field strength and morphology at $L=13,12, \textrm{and}\,11$. The lines show the local orientations of the poloidal magnetic field and the color table denotes the magnetic field strength. Two key features are immediately apparent. Within the disk region $r\lesssim 40$ au and near the midplane, the ohmic dissipation has moved the magnetic field strength peak off center, and it now resides around the edge of the disk. A second key feature is the extremely flared magnetic field lines at $r \gtrsim 40$ au. The poloidal magnetic field lines resemble the lines of a split magnetic monopole in the exterior region of the disk. This is a consequence of rapid infall with near flux freezing. A split-monopole field has long been known as a theoretical outcome occurring at the center of a magnetized collapse with flux freezing \citep[see][]{gal06}. The field inside the disk is peaked toward the disk edge due to ohmic dissipation. The local peak of magnetic field strength leads to some field line segments near the disk surface being directed slightly back towards the $r=0$ axis. The disk edge is the interface between the outwardly diffusing magnetic field inside the disk and the magnetic field outside the disk that is rapidly advected inward. In summary, the magnetic field outside the disk resembles one side of a split-monopole field when viewed from farther radii, while immediately inside the disk there exists a diffused field that peaks at the disk edge.

\begin{figure*}
\plotone{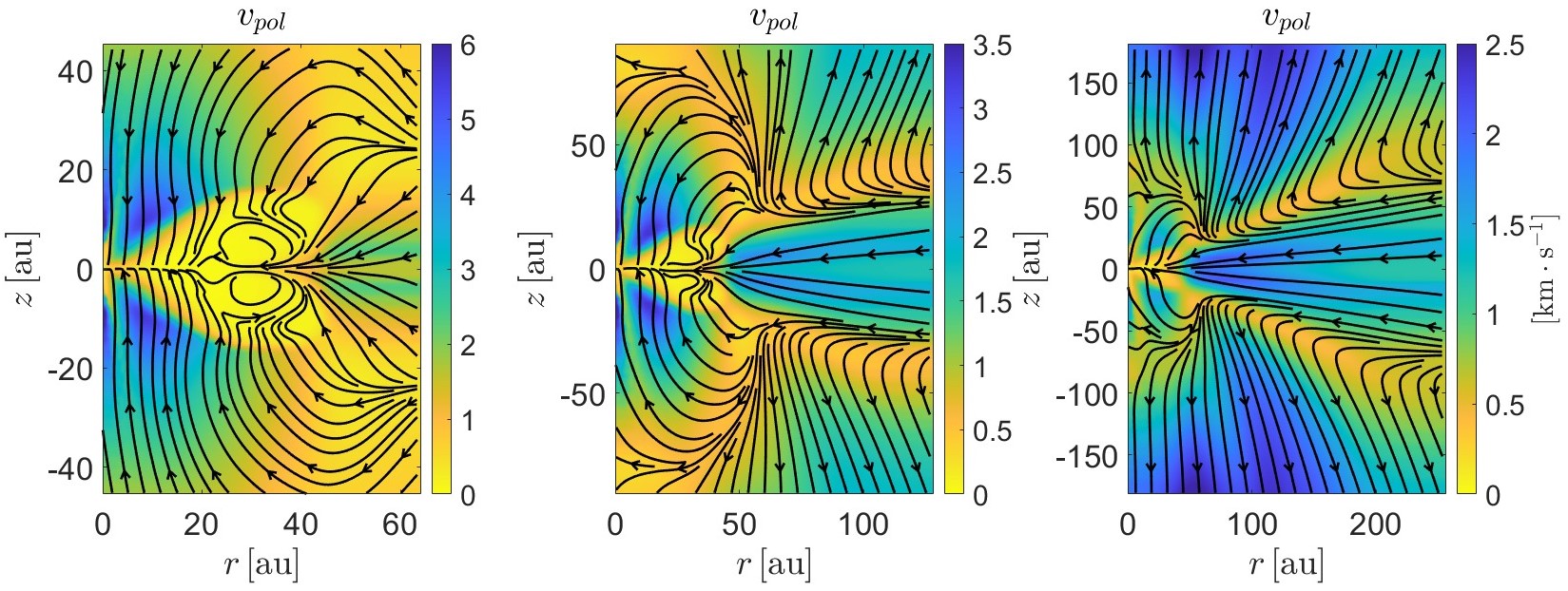}
\caption{Lines of poloidal velocity field $v_{\rm pol} = (v_r^2 + v_z^2)^{1/2}$ at levels $L=$ 13, 12, and 11 (left to right) with the color table representing the field amplitude in km s$^{-1}$. Note the different color scale for each panel.}
\label{fig:vzplots}
\end{figure*}

\begin{figure*}
\plotone{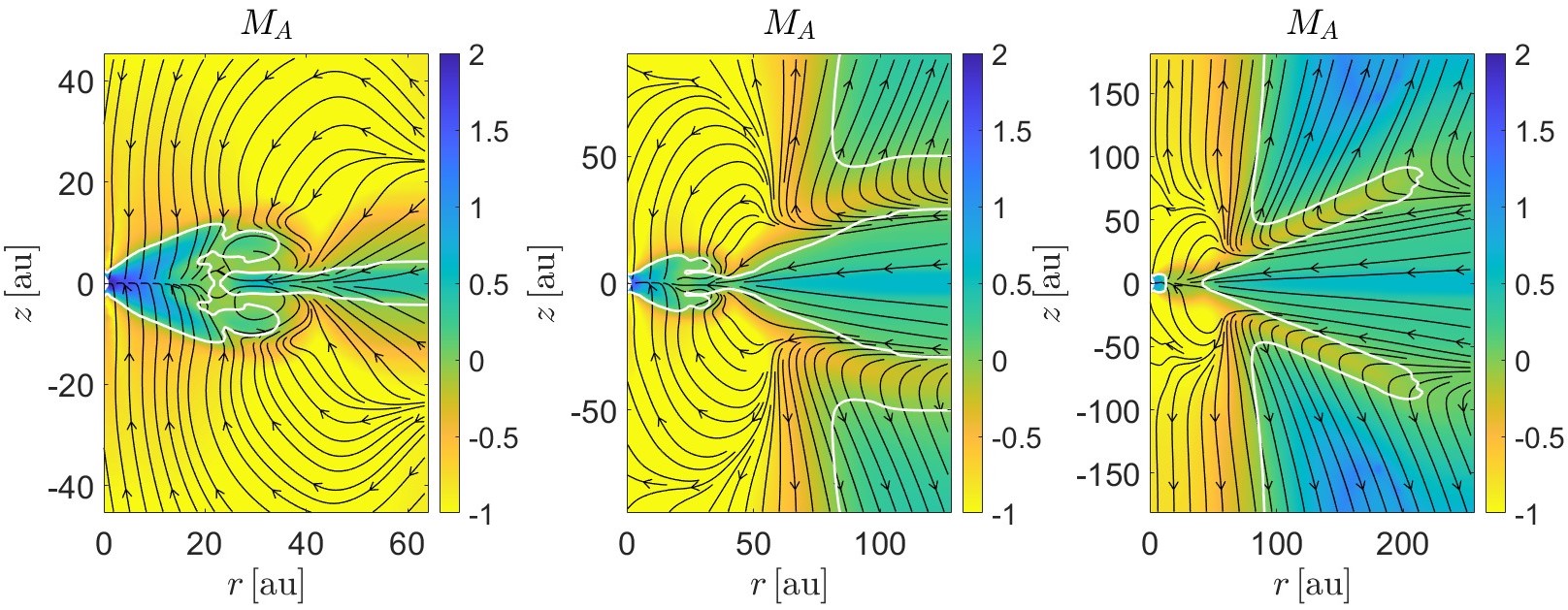}
\caption{Lines of poloidal velocity field at levels $L=13,12,11$ superimposed over an intensity map of the \alfvenic Mach number $M_A \equiv v_{\rm pol}/v_A$ where $v_A$ is the local \alfven speed calculated using the poloidal magnetic field. The white line represents the \alfven surface and the color table is in logarithmic values.}
\label{fig:alfven_plot}
\end{figure*}

Figure~\ref{fig:vzplots} shows the poloidal velocity vectors and corresponding scalar field at the same grid levels as in Figure~\ref{fig:bzplots}. One of the most striking features is that there is no wind emerging from the disk. This is consistent with the degree of magnetic decoupling in the disk that is seen in Figure~\ref{fig:bzplots}. Indeed there is mass infall onto the disk. An outflow is still present in the system but it emerges from the pseudodisk. The strongest outflow (dark blue color in the rightmost panel) emerges from the pseudodisk region just outside the disk, at radii $\simeq 70-100$ au and at a height $z \simeq 30 - 50$ au, which is above the top surface of the disk.  The field lines have a modest inclination to the vertical direction in this launch region, and here the MPGF is driving the flow. Furthermore, at larger radii, there is an outflow that is emerging with relatively lower velocity and in the region of highly inclined field lines above the pseudodisk, shown in yellow color. This flow into a region with highly inclined field lines is a form of MCW even if the initial launch from the pseudodisk can be due to the MPGF. 
In the idealized MCW model of \cite{bla82}, the field lines need to be inclined at least 30$^{\circ}$ outward from the rotation axis for a magnetically-dominated plasma to drive the centrifugal flow. For the MCW launched from the pseudodisk that is not in Keplerian rotation, there is a modification to the inclination criterion. In this case, the MCW flow can be launched from inclination angles less than 30$^\circ$. This agrees qualitatively with the upward flow direction in Figure~\ref{fig:vzplots}, however the many assumptions made in the idealized model limit the ability to compare with the simulation. See Appendix~\ref{ap:mcw} for a derivation of the modified criterion. Even if the pseudodisk is not rotationally supported (as in simplified MCW models), the strongest pseudodisk outflow comes from just outside the disk, where the accretion flow is slowed down by magnetic forces and gas pressure. Furthermore, the outflow is launched from the upper and lower surfaces of the pseudodisk, and not $z=0$ as in standard MCW theory. This means that the magnetic field transports angular momentum from inside the pseudodisk to the surfaces, where the centrifugal effect becomes more important relative to gravity \citep[for a discussion of this process see][]{kud98, tom98}. 

Another special feature of the wind is that the gas from the innermost launch radius $r \simeq 70$ au moves upward due to the MPGF but also inward toward the rotation axis. By the time it reaches a height $z \simeq 50$ au, the gas either continues upward or falls back down onto the disk. In the latter case, gas motion corresponds to a gravity-dominated flow in potential theory \citep[see, for example, fig. 1 in][]{bla82}, as opposed to a magnetocentrifugal flow and the velocity vectors show that this is the case. In the former case, the gas moves slightly inward but is not captured by gravity and continues its upward journey at a smaller radius than the launch point. Both these effects can be seen in the right panel of Figure~\ref{fig:vzplots}. Overall, these principles explain why the region immediately above the disk has a downward velocity but a higher region (above $z \simeq 50$ au in this model) shows an upward flow. This upward flow may appear to be a disk wind if the region below $z \simeq 50$ au cannot be resolved.


Figure~\ref{fig:alfven_plot} illustrates the \alfvenic Mach number $M_A$ overlaid with the poloidal velocity vectors for the same grid levels as Figure~\ref{fig:vzplots}. The \alfven surface, marked by the white line, separates sub-\alfvenic (or magnetically dominated; $\, M_A < 1$) from super-\alfvenic (or kinematically dominated; $\, M_A > 1$) flow. 
The right panel representing $L=11$ particularly shows the complexity of the \alfven surface due to the details of the magnetic field and inflow/outflow velocity field. 
The pseudodisk and disk are kinematically dominated (blue color). 
These regions also build up significant toroidal magnetic field due to their rapid rotation. Hence, a MPGF due mainly to a vertical gradient of the toroidal field at the top and bottom surface of the disk will inevitably launch outgoing winds. There is a region immediately above the disk and part of the pseudodisk that is magnetically dominated (yellow color). In the sub-\alfvenic region above the pseudodisk, the magnetic field lines are also significantly inclined, as explained earlier. This inclination facilitates the outward motion of the wind by the MCW mechanism, a result of the gas being held (at least nearly) in corotation with the dynamically strong magnetic field. However, there is a region further above the pseudodisk that is kinematically dominated (light green color), which has important implications for the outgoing wind. Beyond the \alfven surface, there is a causal disconnection between any traveling hydromagnetic disturbances and influence from the magnetic flux further below. In this super-\alfvenic region, arising due to a weakening of the magnetic field strength at large distance, a twisting of the field occurs as the kinematic motions start to dominate the magnetic field. The MCW mechanism becomes inactive and the MPGF dominates at the larger scales beyond the \alfven surface. For a deeper discussion and review of these effects, see \cite{pud19}.
While we have interpreted the outflow in various regions in terms of the MPGF or the MCW mechanisms, it is worth remembering that each mechanism operates exclusively in one limit: kinematically dominated for the MPGF and magnetically dominated for the MCW. In reality, a hybrid of the two mechanisms is operating in many regions \citep[see also fig. 14 of ][and associated discussion]{mac08}. A very detailed study of the driving mechanism is beyond the scope of the current paper and is left for future work.

\begin{figure}
\plotone{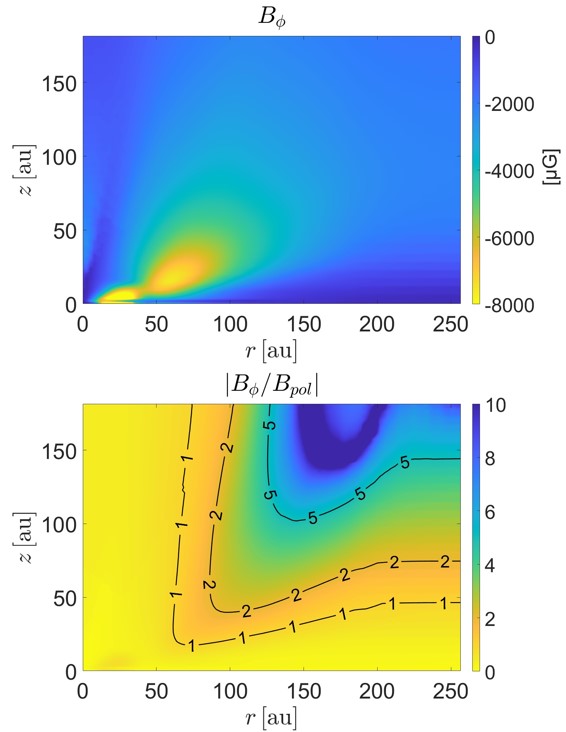}
\caption{Top: toroidal magnetic field component at $L=11$ with the  color table in $\upmu \textrm{G}$. Bottom: ratio $|B_{\phi}/B_{\rm pol}|$ for the same scale and time as the above panel with the level curves for $|B_{\phi}/B_{\rm pol}|=1,2,5$ displayed.}
\label{fig:bphioverbpol}
\end{figure}

Figure~\ref{fig:bphioverbpol} illustrates the toroidal magnetic field component $B_\phi$ in two ways. The top panel is an image of the strength of $B_\phi$ and shows that there is significant twisting of the magnetic field in the kinematically-dominated regions. This occurs inside the disk and separately in a region of the outflowing gas that is beyond the \alfven surface. Since the total magnetic field strength is decreasing with distance from the disk edge, the twisting can be understood better by looking at the bottom panel, which is an image of the ratio $|B_\phi/B_{\rm pol}|$. This panel shows the dramatic increase of this ratio even as the overall field strength is diminishing. The toroidal field begins to dominate when the wind reaches the super-\alfvenic region. Specifically, the contour for $|B_{\phi}/B_{\rm pol}|=2$ coincides well with the contour of the \alfven surface, which can be seen by comparing Figure~\ref{fig:bphioverbpol} with the right panel of Figure~\ref{fig:alfven_plot}.

\begin{figure}
\epsscale{1.1}
\plotone{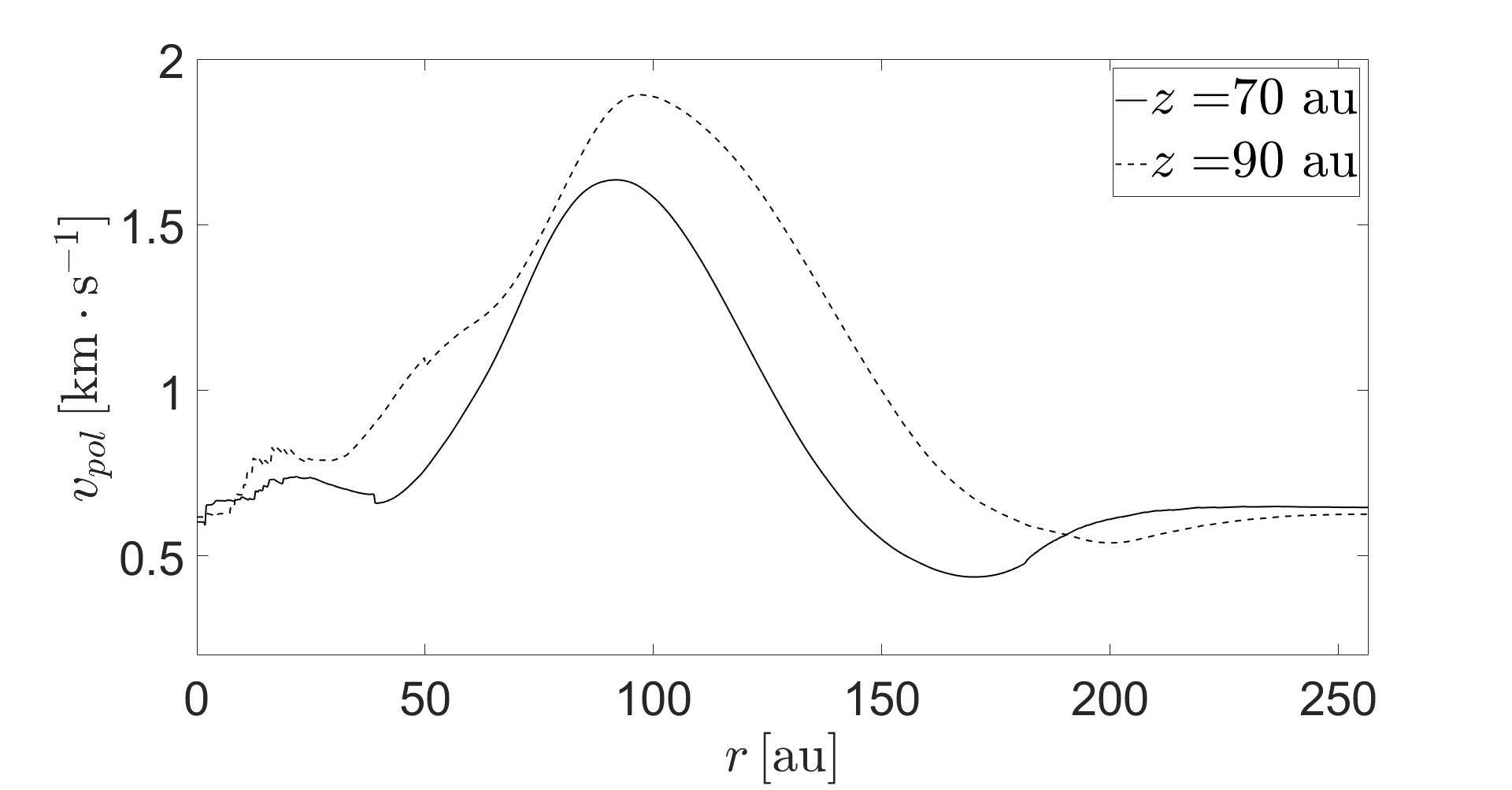}
\plotone{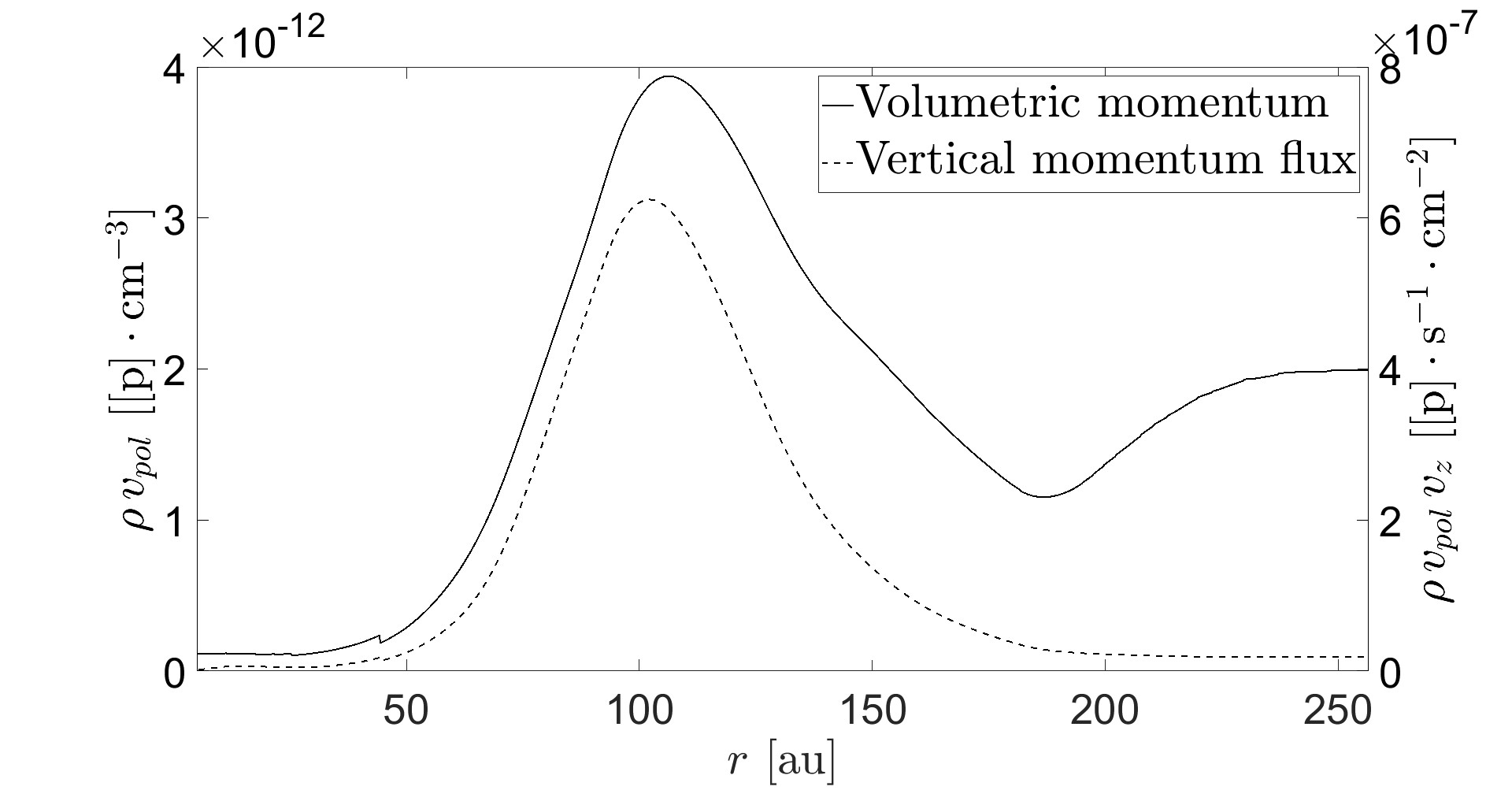}
\caption{Top: poloidal outflow speed $v_{\rm pol} = (v_r^2 + v_z^2)^{1/2}$ versus radius along $z=70$ and $90$ au at level $L=11$. Bottom: poloidal momentum density $\rho \, v_{\rm pol}$ and vertical momentum flux $\rho \, v_{\rm pol}\, v_z$ versus radius along $z=70$ au. Here $[p]$ denotes the unit of momentum, g cm s$^{-1}$.}
\label{fig:Vpol_linegraph}
\end{figure}

Figure~\ref{fig:Vpol_linegraph} shows outflow variables at heights above the disk and pseudodisk where the outflow is newly established. The top panel shows the variation of outgoing speed $\vpol$ versus radius and that there is a clear peak at $r \simeq 100$ au at both heights; this is significantly outside the disk outer radius $r \simeq 40$ au. Clearly the greatest poloidal outflow speeds are from a region above the inner pseudodisk. Note that the outgoing speeds at these heights and at radii $40 \lesssim r \lesssim 150$ au are greater or significantly greater than the outgoing speeds at $r \lesssim 40$ au. The ratio of surface areas of these two regions is $(150^2 - 40^2)/40^2 = 13$. Since the density in the pseudodisk region is lower than in the disk region (see Figures \ref{fig:densityimage} and \ref{fig:densitylinegraphs}), we also consider the density dependent quantities of momentum density and momentum flux, shown in the bottom panel of Figure~\ref{fig:Vpol_linegraph}. These show a strong dominance of both quantities above the pseudodisk region compared to above the disk region. Both quantities peak at $r \simeq 100$ au. The momentum flux resembles a Gaussian function and a best fit of that function yields a peak at $r=96$ au and a standard deviation of $15.6$ au. 

\begin{figure}[!ht]
\plotone{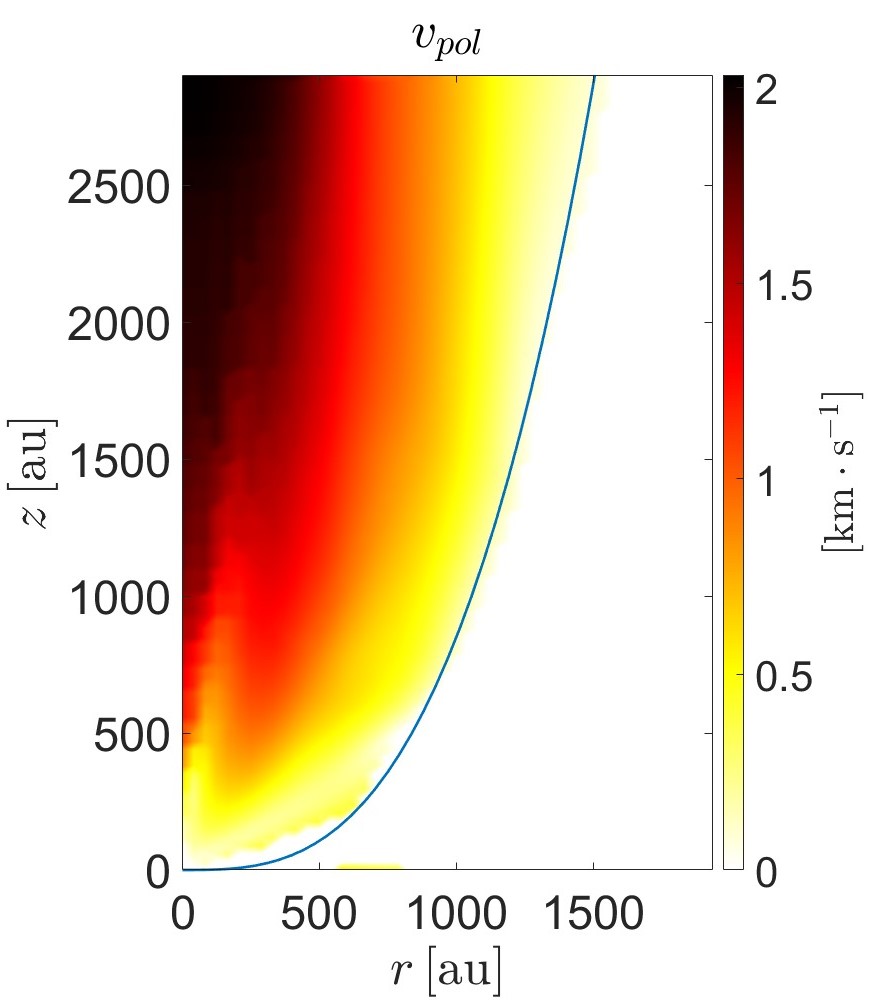}
\caption{Poloidal velocity $v_{\rm pol}$ in km s$^{-1}$ at level $L=7$. Only positive $v_z$ values corresponding to supersonic flow are shown. The boundary of this region, shown in a blue line, is the analytic function $f(r) = \alpha \, r^3$, where $\alpha = 8.5 \times 10^{-7} \, \textrm{au}^{-2}$. The color table represents values in km s$^{-1}$.  }
\label{fig:vpolL7}
\end{figure}

Figure~\ref{fig:vpolL7} shows the outflow zone using a color mapping of the poloidal velocity amplitude of the supersonic outflow. The outflow zone can be fit with the analytic function shown with a blue line for each of $z>0$ and $z<0$. Within each of these surfaces, enclosed physical quantities such as the total outflow mass and momentum can be efficiently calculated. By measuring out to $z=\pm \, 6000$ au, we calculate a total outflowing mass $M_{\rm out}=0.11 \, \msun$, which is $1/3$ of the accreted protostar mass $M_*=0.33\, \msun$ at this time. Using the mean value theorem, a volume integral of the momentum density $\rho\, |v|$ of outflowing material yields a total momentum $p_{\rm out} = 1.34 \times 10^{37}\; \textrm{g \, cm\, s}^{-1} = 6.73 \times 10^{-2}\, \msun\, \textrm{km\, s}^{-1}$.  
This means that the characteristic speed of the outflow is $p_{\rm out}/M_{\rm out} = 0.61$ km s$^{-1}$. We can write as a simple estimate that $p_{\rm out} \simeq 0.1 \,\msun \times 0.6 \, \textrm{km\, s}^{-1} \simeq (1/3) M_* \times 3 c_{\rm s} = M_*c_{\rm s}$, where $c_{\rm s}$ is the isothermal sound speed of the ambient cloud and equal to $0.2$ km s$^{-1}$. 

\begin{figure}
\epsscale{1.1}
\plotone{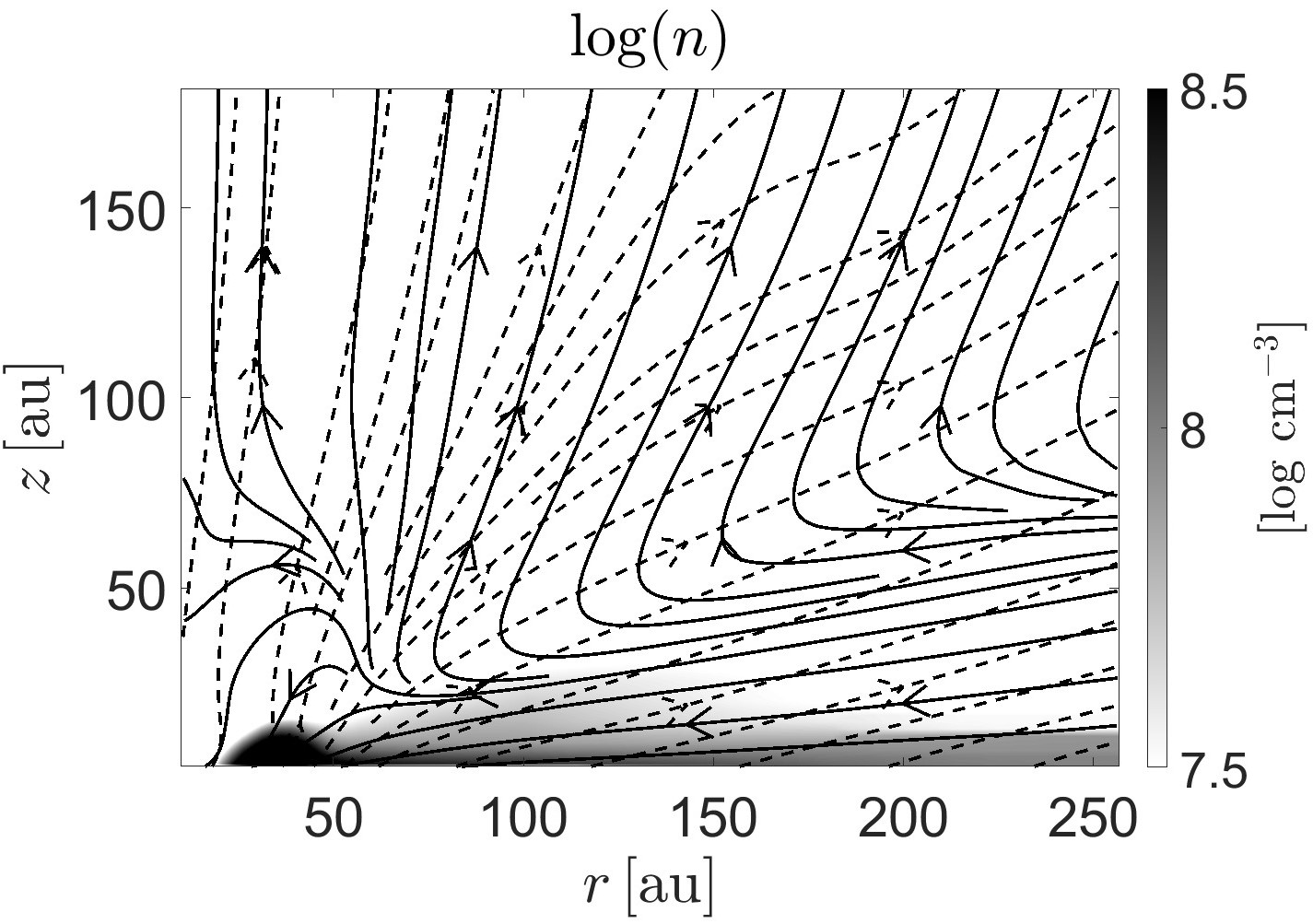}
\caption{Number density in the half-plane $z > 0$ at level $L=11$. The innermost 10 au is omitted to avoid regions influenced by the sink cell. The solid lines represent the poloidal velocity field whereas the dashed lines represent the poloidal magnetic field. The greyscale color table is in logarithmic values and is confined to the range [7.5, 8.5] to highlight the pseudodisk and disk.}
\label{fig:Density_L11}
\end{figure}

Figure \ref{fig:Density_L11} illustrates the logarithmic number density for the half-plane $z>0$ at $L=11$ with two vector field overlays. The solid streamlines correspond to the poloidal velocity field whereas the dashed streamlines correspond to the poloidal magnetic field. The pseudodisk, which is highlighted by the horizontal black strip at the bottom, is characterized by number densities in the range $\sim 10^{7.5} - 10^{8.5}$ cm$^{-3}$. 
In the outflowing region at $r \lesssim 50$ au and above $z \simeq 50$ au, the magnetic and velocity fields are parallel. 
Figure \ref{fig:alfven_plot} shows that this region is also sub-Alfv\'enic, so that the MPGF due to $B_\phi$ drives an outflow along the nearly vertical field lines in the classic manner described by \cite{uch85}. Figure \ref{fig:bphioverbpol} shows that $B_\phi$ and $B_{\rm pol}$ have comparable strength in this region. 
The magnetic field becomes more inclined as one looks toward the right side of Figure \ref{fig:Density_L11}. Here, the alignment between the two fields begins to break down and the wind driving is a more complex effect.
In the sub-\alfvenic region (see Figure~\ref{fig:alfven_plot}) there is a combination of a MCW effect and driving by the MPGF. However, further to the upper right, the flow enters the super-\alfvenic regime and the MPGF mechanism dominates. Figure~\ref{fig:bphioverbpol} shows that $B_{\phi}$ dominates $B_{\rm pol}$ in this region. 



\section{Discussion}
\label{sec:discussion}

The results presented in this paper demonstrate the importance of the pseudodisk as the actual driver of outflows within protostellar cores. The centrifugal disk is usually thought to be the driver of outflows, however a careful analysis of our simulation result shows that there is an extensive pseudodisk region spanning a radius of several hundred au that drives the outflow. The strongest outflow comes from a region that is nearly a factor of two farther in radius than the disk edge. It is the interaction zone between the magnetic flux that is being inwardly advected from the pseudodisk and the magnetic flux diffusing outward through the disk. We call this region the disk interaction zone (DIZ). On its outer side is a highly inclined magnetic field that can resemble the field of a split monopole when viewed from a large distance. On the inner side of the DIZ is a more complex magnetic field that has undergone significant diffusion, with an off-center peak and some segments of the poloidal field lines pointing inward. This distorted structure results in flows being ejected with an inward velocity component. The disk itself has poor coupling with the magnetic field. Therefore, rather than driving an outflow, it is actually receiving the infalling gas that was launched from the inner side of the DIZ. Observational evidence that the outflow launching is occurring from a region somewhat \textit{outside} the disk edge has emerged recently from high-resolution ALMA observations of the Class I object BHB07-11 \citep{alv17}.

Our results establish a new paradigm that winds are actually driven from the pseudodisk, and especially the DIZ, rather than the disk itself. This is a qualitatively new way to think about winds, as the launch region is not in Keplerian rotation. Note that we are talking about the winds here and not an inner jet that is excluded in our model due to having a sink cell of size 2 au. When modeling with sufficient mesh resolution to resolve the central protostar, three-dimensional resistive MHD simulations reveal an inner high-velocity jet as well \citep[e.g.,][]{mac19}. The jet has speed $\sim 10-100$ km s$^{-1}$ and is driven by the MPGF in the innermost region where the high temperature leads to good magnetic coupling of the gas. In terms of overall impact on the ambient molecular cloud, the pseudodisk wind is the more consequential, given its larger launch region and momentum input.
The use of a sink cell allows us to avoid extremely small time steps imposed by the Courant-Friedrichs-Lewy (CFL) condition, and therefore integrate deep into the protostellar phase. At these times the outgoing expansion wave has enclosed a substantial mass and volume; there is a well-developed disk as well as a large region of dynamically-infalling pseudodisk. Furthermore, our nested grid approach allows for good resolution not only in the innermost regions, but also in the intermediate scales of the pseudodisk and the region immediately above it. At the level $L=11$ shown in many of the figures, the cell size is $5.76$ au. The physics that is captured in our simulation may not be seen in simulations that do not integrate to late times and/or maintain good resolution in the intermediate scales. 
An additional problem with the CFL condition is that it imposes severe restrictions on three-dimensional MHD simulations since the motion of matter along field lines evacuates some regions to the point where they have very low density and consequently very high \alfven speed. The maximum \alfven speed in the simulation box determines the time step, which is essentially the minimum Alfv\'en-crossing time across any cell. In our simulations, the minimum plasma $\beta$ is around $10^{-5}$, which means that the \alfven speed is about 300 times greater than the sound speed. We overcome this limitation by using a vector-type supercomputer, the NEC SX-aurora TSUBASA.
Our code is tuned to calculate the simulation on this platform and a typical run presented in this paper takes about 10 weeks of wall clock time.
We did not need to impose any artificial approximations to alleviate the restrictions due to regions of high \alfven speed. 

The pseudodisk is a dynamically infalling flattened structure with a rapid rotation rate in its inner regions. It is important to distinguish the prestellar pseudodisk, which may be a quasistatic structure, from the smaller protostellar pseudodisk that exists within the expansion wave. The latter structure tends to have a very sharp differential rotation $\Omega \propto r^{-2}$ and rapid collapse that drags in the magnetic field much faster than it can be diffused \citep[see][]{dap12}. This results in an extremely pinched magnetic field configuration. The pseudodisk is, however, extremely flattened due to the magnetic pinching force $- \partial/\partial z \,(B_r^2/8 \pi)$ in addition to self-gravity. The radial component $B_r$ is zero at the midplane but rises rapidly in magnitude above and below it, hence the MPGF acts toward the midplane. The disk is less flattened than the pseudodisk due to its higher temperature (see Figure~\ref{fig:densityimage}) and also due to the reduced magnetic pinching force; the field lines contain relatively much less radial component in the disk compared to the pseudodisk, as can be seen in Figure~\ref{fig:bzplots}.

The rapid rotation in the pseudodisk and disk leads to a significant toroidal magnetic field in the regions near the equatorial plane. The gradient in this field is a cause of the initial ejection of the wind. We can consider the ultimate origin of the wind to be the MPGF, however the flow leaving the pseudodisk enters into a magnetically-dominated region with highly inclined field lines, so in this region the MCW mechanism is active. At greater distances along the flow lines, the gas moves into a kinematically-dominated region, as described earlier, and the subsequent driving is by the MPGF mechanism.

A close comparison of Figure~\ref{fig:bzplots} with Figure~\ref{fig:vzplots} reveals that the outgoing flow velocity and the poloidal magnetic field lines are in a similar direction, but are \textit{not} always parallel. There is close alignment for the flow that originates at the DIZ, but at larger radii, the flow emerging from the pseudodisk is not so well aligned with the ambient poloidal magnetic field (see Figure \ref{fig:Density_L11}). This is a curious feature that is not in agreement with many steady-state wind models that have $\bm{v}_{\rm pol} \parallel \bm{B}_{\rm pol}$. However, that feature is put into steady-state models by assumption. \cite{con96} has pointed out that more complex steady-state solutions are possible, and that in general one should not assume $\bm{v}_{\rm pol} \parallel \bm{B}_{\rm pol}$. This is particularly apt in the case where there is a steady inflow or outflow of magnetic flux. 
In the simple thought experiment of infall along a cylindrical radius with flux freezing and a purely vertical magnetic field, $\bm{v}_{\rm pol}$ would be perpendicular to $\bm{B}_{\rm pol}$. 
\cite{con96} further points out that significant radial infall motion in the midplane may give the outflowing material an asymptotic $\bm{v}_{\rm pol}$ that is not exactly parallel to $\bm{B}_{\rm pol}$ \citep[see also][]{con17}. Two-dimensional ideal MHD simulations of outflow generation in collapsing clouds also show that $\bm{v}_{\rm pol}$ and $\bm{B}_{\rm pol}$ are not always parallel \citep{tom98}, although not in the same systematic manner as in our simulation. Note that although the MCW theory assumes that $\bm{v}_{\rm pol}$ and $\bm{B}_{\rm pol}$ are parallel, there is no actual treatment in that model of the forces in the driving region. 
In practice, there is an important role played by the MPGF due to the toroidal field component $B_{\phi}$, and this force need not act parallel to ${\bm B}_{\rm pol}$. Future theoretical work can explore the reasons for the systematic angular difference between $\bm{v}_{\rm pol}$ and $\bm{B}_{\rm pol}$ above the pseudodisk.

A limitation of our model is that the nonideal MHD effect is limited to ohmic dissipation, so that the ambipolar diffusion and Hall terms are not accounted for. Furthermore, the ohmic resistivity $\eta$ is included as parameterized form that is pulled outside the differential operator \citep[see][]{mac07}. Therefore, the spatial gradients in $\eta$ are not fully taken into account. However, inclusion of $\eta$ within the differential operator in similar models \citep[][see their eq. 6]{mac18} has shown that it does not make a strong quantitative difference. An example of a model with $\eta$ inside the differential operator is shown in Appendix~\ref{ap:res}. The inclusion of ambipolar diffusion in future works will establish whether the concentrated accumulation of magnetic flux in the DIZ in our model is maintained and at what quantitative level. Among other three-dimensional nonideal MHD simulations that include ambipolar diffusion, \cite{vay18} stop the calculation immediately after second core formation and do not observe an outflow yet generated, whereas the simulation of \cite{xu2021a} follows the evolution to a similar stage as ours. \cite{xu2021a} employ a spherical coordinate grid with a central sink cell, a barotropic pressure-density relation, and the effects of both ambipolar diffusion and ohmic dissipation. There are some similarities between their results and ours. For example, an off-center peak in the magnetic field strength develops in the outer disk of their model. However, the outflow driving is much weaker than in our model. \cite{xu2021a} state that the mass loss through the outflow is significantly less than the mass accreted from the pseudodisk. Although a quantitative value is not given, the ratio of outflow mass to protostar mass seems much less than the approximate one-third in our model, and the outflow speeds are also small. It is not clear how much of the difference in outflow strength in their model is due to the different numerical code and grid structure employed, specifically a relatively low resolution in the outflow driving zone. The inclusion of ambipolar diffusion and a different initial angular momentum profile in their model can also play a role. Future work can explore the role of all these effects explicitly through the study of a large suite of models and by varying the resolution at the intermediate scales of the outer disk and pseudodisk. 

The scenario established in our work is of a pseudodisk wind (PD-wind) originating mainly at a DIZ lying just beyond the disk, with extremely flared magnetic field lines beyond it, and a different field line structure inside the DIZ that includes some inwardly directed field lines. Outward flow along the field lines outside the DIZ is aided by the magnetocentrifugal effect, while the flow that emerges on the inner side of the DIZ is often pointed inward and is captured in the gravity-dominated flow regime of potential theory. This picture bears a remarkable similarity to the X-wind theory \citep{shu94}, even though the wind launching is occurring just outside the disk instead of just outside the protostar. In the X-wind picture, the inwardly advected magnetic field of the disk sets up an extremely flared magnetic field and its interaction with the stellar (nearly dipolar) magnetic field sets up the interaction at an X-point, and strong outflows are launched from its vicinity. The X-point is located just a few stellar radii from the protostar, where the stellar magnetic field truncates the disk. Some outflowing material from the X-point has an inward velocity component toward the direction of the protostar. This gas is funneled back downward in a gravity-dominated flow \citep[see fig. 4 of][]{shu94}. The physics of the inner protostellar region is not included in our model, which has a sink cell of radius 2 au. An inner jet could be launched from the interior by either an X-wind \citep{shu94} or an MPGF-driven flow \citep{mac08}. Both mechanisms require an expected flux refreezing in the innermost regions due to thermal ionization, and the X-wind model also requires significant radial pinching of the inner magnetic field to develop after flux refreezing.

The PD-wind in our fiducial model has an average speed $\simeq 0.6$ km s$^{-1}$ and peak speed $\simeq 7$ km s$^{-1}$ at height $\simeq 1000$ au. These are somewhat lower than many measurements of low-velocity winds that are $\gtrsim 10$ km s$^{-1}$ \citep[see, e.g.,][]{pas23}. Figure~\ref{fig:parameters} shows that a model with lower $\beta_0$ has a greater maximum outflow speed $\simeq 12$ km s$^{-1}$. 
We keep in mind that observations
are sensitive to fast-moving shells and shock interactions that can arise from the faster jet and outflow emerging from the inner regions where magnetic flux freezing is reestablished. For example, see fig. 4 of \cite{mac19}, which shows both the inner high-velocity flow along with an outer low-velocity flow. 
The results of this paper appeal to one origin of the 
outer low-velocity flow that does not have strong interactions with the outer environment. It does not attempt to explain all velocity components seen in observations.

Our results have introduced a new form of wind driving: the PD-wind. This scenario is complementary to the existing paradigms of driving from an X-point, i.e., the X-wind, and driving from a Keplerian disk, i.e., the D-wind. It is possible that all three can coexist in a physical system. Our simulation that reaches the Class 0 phase does show the development of the PD-wind but not a D-wind. The lack of D-wind is due to the significant magnetic diffusivity within the disk, which weakens the magnetic field strength and coupling. It also straightens the field lines, which does not allow the MCW mechanism to work. However, there is a portion of the PD-wind that moves inward and forms an outgoing wind above the disk at $z \gtrsim 50$ au. This flow could be interpreted observationally as a D-wind if the lower region is not resolved.


\section{Summary}

We used high-resolution three-dimensional nested-grid nonideal MHD simulations to reveal new insights into the basic physics of wind driving. We find the following main results that apply to the early protostellar phase.

\begin{enumerate}
    \item The rapidly infalling and rotating magnetic pseudodisk drives an outflowing PD-wind on radial scales up to several hundred astronomical units from the protostar. No wind is driven from the centrifugally-supported disk.
    
    \item The magnetic field is highly pinched in the pseudodisk region, but is significantly diffused inside the disk. The magnetic field strength peaks at the disk edge.
    
    \item The strongest PD-wind launching region is immediately outside and slightly above the disk. This disk interaction zone (DIZ) is the region of interaction between the inwardly advecting magnetic field of the pseudodisk and the outwardly diffusing magnetic field of the disk.
    
    \item Wind that is launched on the inner side of the DIZ moves inward and either falls back onto the inner disk or keeps moving upward while moving toward the symmetry axis. An interesting consequence is that much of the disk surface experiences a mass inflow rather than an outflow.
    
\end{enumerate}

\begin{acknowledgments}
We thank the anonymous referee for comments that significantly improved the manuscript. We also thank Indrani Das, Takahiro Kudoh, and Kohji Tomisaka for valuable discussions. S.B. acknowledges support from a Discovery Grant from NSERC.
This work was supported by the Japan Society for the Promotion of Science KAKENHI (JP21H00046, JP21K03617: MNM), and NAOJ ALMA Scientific Research grant No. 2022-22B.
\end{acknowledgments}

\appendix

\section{Magnetocentrifugal Launch Angle}
\label{ap:mcw}

The magnetocentrifugal wind theory is based on the idea that the magnetic field dominates energetically in the region above and below the disk, and enforces a corotation of the gas and magnetic field, in addition to constraining poloidal flows to be parallel to the poloidal magnetic field. Thus a fluid element rises from the disk along the direction of the poloidal magnetic field while maintaining the rotation rate it had at the disk surface. Forces other than centrifugal and gravitational (due to a central point mass $M$) are not included, and the magnetic force is only implicitly present as a means to constrain the motion as described above. 

The effective potential is
\begin{equation}
\label{eq:Peff}
    \Peff(r,z) = -\frac{1}{2} \Omzsq r^2 - \frac{GM}{\sqrt{r^2 + z^2}}\,.
\end{equation}
For Keplerian rotation $\Omzsq = \Omksq \equiv GM/r_0^3$, where $r_0$ is the initial radius of the outflow at launch from $z=0$.
For an infalling pseudodisk, the specific angular momentum at the launch point is $j_0 = \Omega_0 r_0^2$. As noted from previous work \citep[e.g.,][]{ter84,sai98, dap12} and our simulation, $j_0$ is spatially uniform in the infalling pseudodisk region that is inside the outgoing expansion wave. We can write the general expression
\begin{equation}
    \Omzsq = \frac{j_0^2}{r_0^4} = \frac{j_0^2}{GMr_0}\frac{GM}{r_0^3}= \gamma \, \Omksq\, ,
\end{equation}
where $\gamma \equiv \rcent/r_0$ and $\rcent = j_0^2/GM$ is the centrifugal radius of the mass shell at $r=r_0$.
Next, transform variables to $(r,\theta$), where $\tan \theta = z/r$, to yield

\begin{equation}
    \Peff(r,\theta) = -\frac{1}{2} \gamma\, \Omksq r^2 - \frac{GM}{r \sqrt{1 + \tan^2 \theta}}\,.
\end{equation}
If the rotation is not Keplerian it means that the gas element initially at $r=r_0,z=0$ is not in an equilibrium between gravity and the centrifugal force. In our analysis we assume there are additional forces that keep the initial state in equilibrium. These would be magnetic and thermal forces that reduce the acceleration (formally to zero) of a fluid element in the pseudodisk before it is launched outward. Assuming equilibrium initially, we can study the cases where motion along a direction given by $\theta$ is either stable ($\partial^2 \Peff/\partial r^2 > 0$) or unstable 
($\partial^2 \Peff/\partial r^2 < 0$). First we note that
\begin{eqnarray}
    \frac{\partial \Peff}{\partial r} & = & - \gamma\, \Omksq r + \frac{GM}{r^2 \sqrt{1 + \tan^2 \theta}}\,,\\
    \frac{\partial^2 \Peff}{\partial r^2} & = & - \gamma\, \Omksq - \frac{2GM}{r^3 \sqrt{1 + \tan^2 \theta}}\,.
\end{eqnarray}
The critical angle $\theta_c$ can be found by setting $\partial^2 \Peff/\partial r^2 = 0$ at $r=r_0,z=0$. This leads to 
\begin{equation}
    \tan \theta_c = \pm \sqrt{4 \gamma^{-2} - 1}  \, .
\end{equation}
If $\gamma = 1$ (Keplerian rotation) we get $\tan \theta_c = \pm \sqrt{3}$, and the positive root is $\theta_c = \pi/3 = 60^\circ$. For angles less than $\theta_c$ relative to the equatorial plane, the flow is unstable and a fluid element will keep moving outward.
In the case of pseudodisk launching with $\gamma = 1/2$, i.e., $r_0 = 2 \rcent$, we find that $\tan \theta_c = \pm \sqrt{15}$, and the positive root yields 
$\theta_c = 75.5^\circ$.

The magnetocentrifugal effect in this idealized example can work at a larger angle $\theta$ relative to the equatorial plane for flow above a pseudodisk than for a flow above a Keplerian disk. 
For a given value of $\Omega_0$, the gravitational effect is weaker than in the standard magnetocentrifugal theory when $r>\rcent$. This allows the centrifugal effect to fling a gas element outward at some angles $\theta > 60^\circ$.

\section{Resistivity in Induction Equation}
\label{ap:res}

We have run a model presented in Section~\ref{sec:results} using the form of the induction equation given by Equation~(\ref{eq:induction}). Here we present a comparison of the model B01 with an equivalent model in all aspects except that the right-hand side of Equation~(\ref{eq:induction}) has the correct term $\nabla \times (-\eta\, \nabla \times \bm{B})$ instead of $\eta\, \nabla^2 \bm{B}$. 
The latter term does not fully incorporate the effects of the spatial gradients in $\eta$, however they are partially included since $\eta$ has a density and temperature dependence \citep[see][]{mac07,mac08} and these quantities have a gradient in the region of the protostellar disk (see Figures~\ref{fig:densityimage} and \ref{fig:densitylinegraphs}). 

Given the time required to run simulations and to perform the analysis that is presented in Section~\ref{sec:results}, we do not redo each model. Here, we instead present a brief comparison of model B01 with a revised version. 

\begin{figure*}[ht!]
\begin{center}
\includegraphics[width=1.0\linewidth]{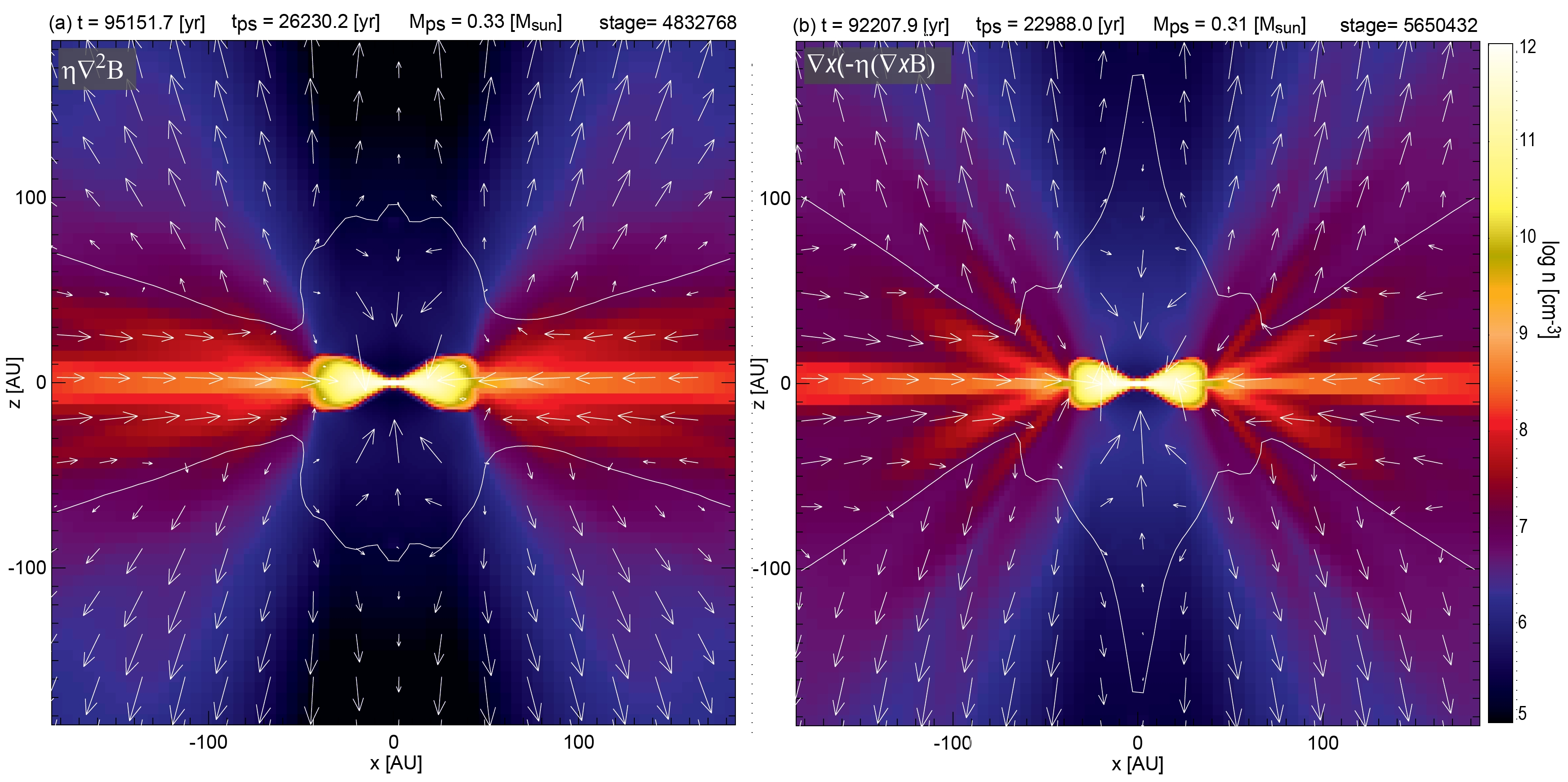}
\end{center}
\caption{Cuts along the $x-z$ plane of the number density with overlaid velocity vectors for model B01 (left) and another model with a revised treatment of resistivity (right; see explanation in text). The normalization of velocity vectors is the same as in Figure~\ref{fig:densityimage}. The two models are shown at approximately the same stage of growth of central mass $M_*$.
At left: $M_*=0.33\msun,\,t=95.2\;{\rm kyr},\,t_{\rm ps}=26.2\;{\rm kyr}$. At right: $M_*=0.31\msun,\,t=92.2\;{\rm kyr},\,t_{\rm ps}=23.0\;{\rm kyr}$.
}
\label{fig:etanabla}
\end{figure*}

Figure~\ref{fig:etanabla} shows images of the number density at level $L=13$ overlaid with velocity vectors for model B01 (left; this is similar to the top left panel of Figure~\ref{fig:densityimage}) and the revised model (right) at essentially the same evolutionary stage i.e., central protostellar mass. The two models show broadly similar features at this stage. The disk size and density distribution are similar, with small but noticeable differences. The outflow structure and velocity magnitudes are very similar, and the key feature of infall directly above the disk and outflow at larger heights is confirmed in the revised model. We leave a detailed examination of the effect of the revised resistivity treatment to a future paper that can explore this as well as other nonideal MHD effects like ambipolar diffusion.

\bibliography{manuscript}{}
\bibliographystyle{aasjournal}

\end{document}